\documentclass{imanum}

\usepackage{color}

\gridframe{N}
\jno{dri017}
\usepackage{modamsthm,modnatbib}
\usepackage{latexsym,amssymb,lastpage}
\usepackage{graphicx,amsfonts}
\usepackage{times,mathptmx,bm,amsmath}
\usepackage{dcolumn}
 \newtheoremstyle{theorem}{6pt}{6pt}{\rm}{}{\sffamily}{ }{ }{}
 \theoremstyle{theorem}

  \newtheoremstyle{thm}{6pt}{6pt}{\rm}{}{\sffamily}{ }{ }{}
 \theoremstyle{thm}

 \newtheoremstyle{lemma}{6pt}{6pt}{\rm}{}{\sffamily}{ }{ }{}
 \theoremstyle{lemma}
 
 \newtheoremstyle{lem}{6pt}{6pt}{\rm}{}{\sffamily}{ }{ }{}
 \theoremstyle{lem}

\newtheoremstyle{case}{6pt}{6pt}{\rm}{}{}{. }{ }{}
 \theoremstyle{case}

 \newtheoremstyle{statement}{6pt}{6pt}{\rm}{}{\sffamily}{ }{ }{}
\theoremstyle{statement}

 \newtheoremstyle{corollary}{6pt}{6pt}{\rm}{}{\sffamily}{ }{ }{}
 \theoremstyle{corollary}

  \newtheoremstyle{defi}{6pt}{6pt}{\rm}{}{\sffamily}{ }{ }{}
 \theoremstyle{defi}

  \newtheoremstyle{cor}{6pt}{6pt}{\rm}{}{\sffamily}{ }{ }{}
 \theoremstyle{cor}

\newtheoremstyle{example}{6pt}{6pt}{\rm}{}{\sffamily}{ }{ }{}
\theoremstyle{example}

\newtheoremstyle{remark}{6pt}{6pt}{\rm}{}{\sffamily}{ }{ }{}
\theoremstyle{remark}

\newtheoremstyle{approximation}{6pt}{6pt}{\rm}{}{\sffamily}{ }{ }{}
\theoremstyle{approximation}

\newtheoremstyle{scheme}{6pt}{6pt}{\rm}{}{\sffamily}{ }{ }{}
\theoremstyle{scheme}

\newtheoremstyle{Algorithm}{6pt}{6pt}{\rm}{}{\sffamily}{ }{ }{}
\theoremstyle{Algorithm}

 \newtheoremstyle{Remark}{6pt}{6pt}{\rm}{}{\sffamily}{ }{ }{}
 \theoremstyle{Remark}

\newtheoremstyle{Lemma}{6pt}{6pt}{\rm}{}{\sffamily}{ }{ }{}
\theoremstyle{Lemma}

\newtheoremstyle{Assumption}{6pt}{6pt}{\rm}{}{\sffamily}{ }{ }{}
\theoremstyle{Assumption}

\newtheoremstyle{Proposition}{6pt}{6pt}{\rm}{}{\sffamily}{ }{ }{}
\theoremstyle{Proposition}

\newtheoremstyle{prop}{6pt}{6pt}{\rm}{}{\sffamily}{ }{ }{}
\theoremstyle{prop}

\newtheoremstyle{rem}{6pt}{6pt}{\rm}{}{\sffamily}{ }{ }{}
 \theoremstyle{rem}

\newtheoremstyle{hypo}{6pt}{6pt}{\rm}{}{\sffamily}{ }{ }{}
 \theoremstyle{hypo}

  \newtheoremstyle{Step}{6pt}{6pt}{\rm}{}{}{ }{ }{}
 \theoremstyle{Step}

 \newtheoremstyle{lema}{6pt}{6pt}{\rm}{}{\sffamily}{ }{ }{}
 \theoremstyle{lema}

\newcommand{\la}{\langle}
\newcommand{\ra}{\rangle}

\newcommand{\ds}{\displaystyle}
\newcommand{\avh}{\la H_1 \ra}
\newcommand{\pa}{\partial}
\newcommand{\ud}{{\mathrm d}}
\newcommand{\ep}{\varepsilon}
\newcommand{\w}{\omega}
\renewcommand{\theequation}{\thesection.\arabic{equation}}
\numberwithin{equation}{section}
\def\citeasnoun{\cite}

\begin{document}
%%%%%%%%%%%%%%% this is title page style%%%%%%%%%
\title{Nonlinear Waves in Lattices: Past, Present, Future}
\author{{\sc P.G. Kevrekidis}\\[2pt]
Department of Mathematics and Statistics, University of Massachusetts, \\[6pt]
 Amherst, MA 01003, USA.\\[6pt]}
%{\rm [Received on 18 February 2005]}}
\pagestyle{headings}
\markboth{P.G. Kevrekidis}{\rm Nonlinear Waves in Lattices: Past, Present, Future}
\maketitle
%%%%%%%%%%%%%%%%%

\begin{abstract}
{In the present work, we attempt a brief summary of various
areas where nonlinear waves have been emerging in the phenomenology
of lattice dynamical systems. These areas include nonlinear optics,
atomic physics, mechanical systems, electrical lattices, nonlinear
metamaterials, plasma dynamics and granular crystals. We give some
of the recent developments in each one of these areas and 
speculate on some of the potentially interesting directions
for future study.}

\end{abstract}
%%%%%%%%%%%%%%%%%%%%%%%%%%%%%%%

\section{Introduction}

Over the last two decades, there has been an explosion of interest
on the dynamics of nonlinear waves in lattices. It can be argued
that this field was actually initiated by the seminal investigation
of Fermi, Pasta and Ulam (FPU) presented in
\citeasnoun{fpu54}, who posed the following
question:
How long does it take for long-wavelength oscillations to transfer
their energy into an equilibrium distribution in a one-dimensional
string of nonlinearly interacting particles? This question
spurred the activity of a wide array of fields including
soliton theory, discrete lattice dynamics and KAM theory,
all of which remain active research fields today.

The development of soliton theory, while focusing at the continuum
level at first through the investigations of \citeasnoun{zk67}
and the concomitant developments of the theory of integrability for the
Korteweg-de Vries equation, also offered some of the first
particularly interesting examples of lattice nonlinear dynamical
systems. Among the prime examples thereof, one can classify the
Toda lattice of \citeasnoun{toda81}, the Ablowitz-Ladik equation of
\citeasnoun{AL76} and the Calogero-Moser $N$-body problem, see e.g.
the works of \citeasnoun{cal} and \citeasnoun{mos}.

It was at around the same time that some of the early fundamental
suggestions of the role of discrete solitons in physical systems
were made in the 1970s. Some of the pioneers of these contributions
were Davydov, see e.g. \citeasnoun{dav72} and Heeger and collaborators,
see e.g. \citeasnoun{Su} and \citeasnoun{Su1}.
Davydov proposed the discrete soliton as
a tool for understanding energy transfer in proteins and their
$\alpha$-helices, while Heeger and collaborators put forth solitonic
models for understanding neutral and charge transport in conducting
polymer chains such as polyacetylene and polythiophene.

However, a true explosion of the field came about in the late
1980s when not only major developments arose in the context
of the so-called discrete self-trapping equation of \citeasnoun{dst},
but also the work of \citeasnoun{st88} and \citeasnoun{skt88}
brought forth the theme of intrinsic localized modes (or discrete
breathers) in anharmonic lattices. Although this theme had been
explored earlier e.g., by
\citeasnoun{ovch69}, \citeasnoun{koskov74} and \citeasnoun{dolgov86},
it was \citeasnoun{st88} and \citeasnoun{skt88}
that motivated a number of other researchers to explore the
existence and stability of such modes more thoroughly and to
start obtaining fundamental results about their properties,
as well as to devise special limits (such as the anti-continuum (AC)
limit of lattices with uncoupled sites) which could be used to
showcase the
generic nature of these modes. A landmark in connection to these
efforts was the work of \citeasnoun{macaub} which
established rigorously the existence/robustness of such localized modes
starting from the AC limit, under minimal appropriate
(non-resonance) assumptions for chains of nonlinear oscillators.

Since these early steps, there has been a tremendous amount
of developments in the area of nonlinear waves in dynamical lattices.
This activity has chiefly been fueled by the experimental
observation of these modes in a wide range of physical systems
in biophysics, solid state physics, nonlinear optics, atomic
physics, granular crystals, and plasma physics, in addition
to numerous developments in the more classical fields of
mechanical and electrical lattices [and even the latter have seen
interesting recent developments e.g. with the emergence of
the field of nonlinear metamaterials, among others, as discussed
in section \ref{future1} below]. A (partial) list of
key physical systems where explanations in terms of
nonlinear waves in lattices have had an impact includes:
\begin{itemize}
\item The observation of discrete breathers in
complex electronic materials such as halide-bridged transition metal
complexes as e.g. in \citeasnoun{swanson}.
\item The formation of denaturation bubbles in the DNA double
strand dynamics summarized e.g. in \citeasnoun{peyrard}.
\item The emergence of dynamical instabilities, discrete quantum
self-trapping and localized modes in ultracold Bose-Einstein
condensates in the presence of deep optical lattices, see e.g. the reviews of
\citeasnoun{mplbkon} and \citeasnoun{oberthaler}.
\item The illustration of localized modes of solitonic, vortex,
ring, multipole, surface, gap, necklace and numerous other types
in the nonlinear optics of evanescently coupled waveguide arrays,
as well as in that of biased photorefractive crystals;
a relevant recent review can be found in \citeasnoun{discreteopt}.
\item The prototypical mechanical
realization of solitary waves and localized
modes in driven micromechanical cantilever arrays as shown
in \citeasnoun{sievers},
as well as in coupled torsion pendula where a recent example
is given by \citeasnoun{lars}.
\item The examination of interesting properties and mobility
of localized modes in nonlinear electrical lattices and
transmission lines; see e.g. \citeasnoun{rem} and \citeasnoun{lars2}.
\item The presence of nearly compact solitary waves in granular
crystals consisting of beads with Hertzian elastic
interactions reviewed in \citeasnoun{nesterenko1} and \citeasnoun{sen}.
\item The creation of discrete breather type structures in layered
antiferromagnetic samples such as those of a (C$_2$H$_5$NH$_3$)$_2$CuCl$_4$
crystal, upon imposition of a dynamically unstable continuous wave
(which falls into modulational instability and results in the
localized waveforms); see the relevant works of
\citeasnoun{lars3} and \citeasnoun{lars4}.
\end{itemize}

These experimental developments have, in turn, motivated
a huge range of theoretical advances in connection to the
properties and interactions of these lattice nonlinear waves
and on how these are modified in comparison to their continuum
siblings. In particular, some of the areas that have seen intense
theoretical activity concern:
\begin{itemize}
\item The existence and stability of the localized modes.
\item The dynamics and interactions of multiple such states in nonlinear
lattices.
\item The effect of external potentials (either local ``impurity-like''
potentials, or more broad externally imposed, field potentials)
\item The effect of short-range (local) versus long-range (nonlocal)
interactions.
\item The presence of excitation thresholds for such nonlinear states
(especially in higher dimensional settings).
\item The statistical mechanics and long-time asymptotics of these
lattice systems and the role of the localized modes in them.
\item The effects of disorder and the role of the interplay between
the linear phenomenon of Anderson localization and of the nonlinearity-induced
type of localization.
\item The interaction with and scattering of phonon-like modes from
discrete solitary waves.
\item The modifications imparted on the waveforms of such localized states and
the new possibilities for ones such (e.g. with embedded vorticity)
afforded by higher dimensional settings.
\end{itemize}
Many of these topics have been summarized in a number of
reviews by \citeasnoun{braun}, \citeasnoun{aubry1}, \citeasnoun{flach1},
\citeasnoun{hennig}, \citeasnoun{KRB01}, \citeasnoun{eilbeckjoh}, \citeasnoun{flach2}
and more recently in a number of books e.g. about Klein-Gordon
nonlinear lattices by \citeasnoun{braunbook}, as well as about
nonlinear Schr{\"o}dinger ones by \citeasnoun{pgkbook}.

As it can be inferred from the shear volume of review articles
on the topic (as well as that of full-blown books), it is impossible
to give an exhaustive description of any particular aspect
or application of the subject in anything less than a lengthy
review or a book. Our aim herein will thus, by necessity, be more
modest, but in some sense, also more ``forward looking''.
We aim to give a selective presentation of some of recent
developments of a few among these areas, which in the present
author's evaluation hold significant potential for
future developments in the field of localized modes, nonlinear
waves and discrete breathers. Obviously, this bears a significant
weight of subjective perception and a personalized viewpoint about
which we caution the reader in advance. On the other hand, our
hope is to offer a number of interesting directions for future
study in these areas which could significantly enhance our
understanding of nonlinear waves in these settings, promoting
their state-of-the-art and offering significant connections with
present or near-future experimental settings and potentially also with
applications.

We will categorize our presentation based on the
different models and the corresponding areas of interest. Hence, upon
introducing our basic notation,
the
third section will concern itself with localized modes
in models of the discrete nonlinear Schr{\"o}dinger
type and their applications
in nonlinear optics and atomic physics. The fourth section will
consider models of the Klein-Gordon type, as well as their
mechanical and electrical (as well as nonlinear metamaterial)
applications. The fifth section
will focus on
% regular but also generalized forms of Klein-Gordon
%models as they arise in the study of electrical lattices,
%nonlinear metamaterials, as well as in the modeling of dusty plasmas.
%The sixth section will examine
models of the FPU type
arising in the examination of granular crystals with Hertzian
(or modified Hertzian) interactions, as well as in the case
of study of dusty plasmas. Lastly, in section six,
we will summarize some of our discussion and offer some
concluding thoughts.

\section{Notation}

The prototypical dynamical lattices that will be considered
in this review will be the discrete nonlinear Schr{\"o}dinger
equation (DNLS) in the form:
\begin{equation}
i \dot{u}_n=-\epsilon (u_{n+1}+u_{n-1}) - |u_n|^2 u_n,
\label{s5_45}
\end{equation}
the nonlinear Klein-Gordon (KG) lattice of the form:
\begin{eqnarray}
\ddot{u}_n= \epsilon (u_{n+1}+u_{n-1}-2 u_n) - V'(u_n),
\label{rev_eq2}
\end{eqnarray}
and the FPU-type dynamical lattice of the form
\begin{eqnarray}
\ddot{u}_n = V'(u_{n+1}-u_n) - V'(u_{n}-u_{n-1}),
\label{rev_eq3}
\end{eqnarray}
where $u_n$ is the dynamical variable of interest measured
at site $n$, $\epsilon$ is the coupling parameter and $V$
a potential (that could be onsite for KG lattices or inter-site
for FPU ones).

It should be noted that in what follows, we will not restrict
ourselves to the one-dimensional framework of Eqs. 
(\ref{s5_45})-(\ref{rev_eq3}), but motivated by physical
applications, we will often consider higher dimensional
variants of the models. In their prototypical form, two-dimensional
generalizations of Eqs. (\ref{s5_45})-(\ref{rev_eq3}) read
\begin{equation}
i \dot{u}_{n,m}=-\epsilon (u_{n+1,m}+u_{n-1,m}+u_{n,m+1}+u_{n,m-1}) - 
|u_{n,m}|^2 u_{n,m},
\label{s5_new1}
\end{equation}
for the DNLS lattice, 
\begin{eqnarray}
\ddot{u}_{n,m}= \epsilon (u_{n+1,m}+u_{n-1,m}+u_{n,m+1}+u_{n,m-1}-4 u_{n,m}) - V'(u_{n,m}),
\label{s5_new2}
\end{eqnarray}
for the KG case, while for the FPU-type lattices we have
\begin{eqnarray}
\ddot{u}_{n,m} = V'(u_{n+1,m}-u_{n,m}) - V'(u_{n,m}-u_{n-1,m})
+ V'(u_{n,m+1}-u_{n,m}) - V'(u_{n,m}-u_{n,m-1}).
\label{s5_new3}
\end{eqnarray}
This, in turn, clearly suggests how to expand the model beyond
two dimensions and into the three- or arbitrary-dimensional-case, by
sequentially adding more lattice directions and linear (in cases
such as (\ref{s5_new1})-(\ref{s5_new2})) or nonlinear (as in case
(\ref{s5_new3})) interactions along them.

In the first one of these, the field will be complex
and we will most often seek stationary (standing wave)
type solutions of the form $u_n=\exp(i \mu t) v_n$; $\mu$ is
often referred to as the propagation constant in nonlinear
optics, or the chemical potential in Bose-Einstein condensates
and characterizes the frequency of such standing wave
solutions. These
will satisfy the steady state equation:
\begin{equation}
(\mu-|v_n|^2) v_n = \epsilon (v_{n+1}+v_{n-1}).
\label{s5_46}
\end{equation}
Additionally, one of the fundamental ideas that we will
use in exploring this class of systems will be that
of the anti-continuum limit of \citeasnoun{macaub}.
This is the limit where the sites are uncoupled with
$\epsilon=0$. In this case, Eq.~(\ref{s5_46})
is completely solvable $v_n=0,\pm \sqrt{\mu} \exp(i \theta_n)$,
where $\theta_n$ is a free phase parameter for each site.
The fundamental question of interest will involve
the persistence of these solutions for $\epsilon \neq 0$,
as well as their stability. The latter will be often explored
by means of linear stability analysis, upon imposition of
a perturbation of the form:
\begin{eqnarray}
u_n(t)=\exp(i \mu t) \left[ v_n + \delta \left(p_n \exp(\lambda t)
+ q_n^{\star} \exp(\lambda^{\star} t) \right) \right].
\label{stab_dnls}
\end{eqnarray}
[Notice that the $^{\star}$ will be generally used in what
follows to denote complex conjugation.]
Consideration of the resulting equation upon substitution of (\ref{stab_dnls})
to (\ref{s5_45}), at
${\cal O}(\delta)$ yields the linearization (infinite-dimensional matrix)
eigenvalue problem for the eigenvalues $\lambda$ and eigenvectors
$(p_n,q_n)$.

In the case of the Klein-Gordon models, we will 
%briefly consider  stationary
%solutions, but also more importantly 
focus on discrete breather time-periodic
solutions for different potentials which are either polynomials
in the form $V(u_n)=\sum_{n=2}^{N} a_j u_n^{j}$ or other functions
such as $V(u_n)=1-\cos(u_n)$ (in the sine-Gordon case).
Notice also that we will often use the operator notation
$\Delta_2 u_n = u_{n+1} + u_{n-1} - 2 u_n$ for the discrete Laplacian.
For the FPU
lattices, we will primarily concern ourselves with power law potentials
and particularly the Hertzian elastic case of $V(r_n) \sim r_n^{5/2}$,
but we will also briefly refer to other potentials such as the Yukawa
one $V(r_n) \sim \exp(-\alpha r_n)/r_n$ (for $r_n>0$) in the context
of dusty plasmas. Notice that in the FPU-type chains, we will also
refer to the so-called strain formulation which introduces the modified
field variable $r_n=u_{n-1}-u_n$ and leads to the dynamical equation
\begin{eqnarray}
\ddot{r}_n=V'(r_{n+1})+V'(r_{n-1})-2 V'(r_n).
\label{strain}
\end{eqnarray}

\section{DNLS Models in Nonlinear Optics and Atomic Physics}

\subsection{Recent Developments}

Many of the investigations in these fields stemmed from the
relevance of the DNLS model and its variants in the
nonlinear optics of fabricated AlGaAs
waveguide arrays as in \citeasnoun{7}, as well as in that
of Bose-Einstein condensates in sufficiently deep optical
lattices analyzed in \citeasnoun{mplbkon}, \citeasnoun{oberthaler},
and \citeasnoun{emergent}.

In the former area, a multiplicity
of phenomena such as discrete diffraction, Peierls barriers
(the energetic barrier that a wave needs to overcome to move
from one lattice site to the next),
diffraction management (the periodic alternation of the
diffraction coefficient) in \citeasnoun{7a} and gap solitons
(structures localized due to nonlinearity in the gap of the underlying
linear spectrum) in \citeasnoun{7b} among
others [see also \citeasnoun{eis3}] were experimentally observed.
More recently, fundamental investigations also on the modulational
instability arising in such settings for simple plane wave solutions
in \citeasnoun{MI_dnc}, and on the role of multiple components and
possible four-wave-mixing effects arising from their coupling
in \citeasnoun{2c_dnc} have been explored, as well as that of interactions
of solitary waves with surfaces in \citeasnoun{dnc_surf}.

A related area where lattice dynamical models, although not directly
relevant as the most 
appropriate description (since the setting is, in principle
continuum with a periodic potential),
still yield accurate predictions regarding the existence and
the stability of nonlinear localized modes is that of optically induced
lattices in photorefractive media such as Strontium Barium Niobate
(SBN). Since the theoretical suggestion of such lattices
in \citeasnoun{efrem}, and their experimental realization in
\citeasnoun{moti0}, \citeasnoun{neshevol03} and \citeasnoun{martinprl04},
there has been a tremendous growth in the
area of nonlinear waves and solitons in such periodic,
predominantly two-dimensional, lattices.
The continuously growing array of structures that have been predicted and
experimentally obtained in such lattices includes
(but is not limited to)  various discrete multi-pulse patterns,
such as the discrete dipole solitons in \citeasnoun{dip},
quadrupole solitons in \citeasnoun{quad}, necklace solitons 
in \citeasnoun{neck},
soliton stripes in \citeasnoun{multi},
%impurity
%solitons \citeasnoun{fedele},
discrete vortices in
\citeasnoun{vortex}, and \citeasnoun{vortex1},
rotary solitons in \citeasnoun{rings}, surface solitons in
\citeasnoun{chendnc}, and so on.
%, and others \citeasnoun{us_yuri}.
Such structures have a definite potential to be used as carriers and
conduits for data transmission and processing, in the setting of all-optical
communication schemes; see a detailed recent review in
\citeasnoun{discreteopt}.

Lastly, a completely independent and entirely different field where
relevant considerations and structures, as well as models of the DNLS
type arise is that of ultracold atomic gases in the form of
Bose-Einstein condensates (BECs) under the confinement imposed
by so-called optical lattice potentials. These are periodic
traps produced by counter-propagating
laser beams in one, two or even all three directions, see e.g.
\citeasnoun{bloch}.
This field has also experienced a tremendous amount of development
in the past few years with some of its landmarks being the
prediction [in \citeasnoun{smerzius}] and manifestation
[in \citeasnoun{fallani}]
of modulational instabilities, the observation of gap solitons
in \citeasnoun{markus}, of Landau-Zener
tunneling (tunneling between different bands of the periodic potential)
in \citeasnoun{arimondo} and Bloch oscillations (for matter waves subject
to combined periodic and linear potentials) in \citeasnoun{bpa_kasevich}
among many other salient features.

All of the above settings are described at some level of approximation
by a continuum nonlinear Schr{\"o}dinger (NLS) equation with a periodic
potential. However, when the periodic potential is sufficiently
deep, it was realized that an equivalent description of the system
can be given by a nonlinear dynamical lattice as discussed e.g. in
\citeasnoun{tromb}, and
\citeasnoun{konot}. This was further elaborated for BEC
settings in \citeasnoun{wann1}
[see also \citeasnoun{abl_book} for a related discussion in connection
to nonlinear optics], whose presentation we briefly follow here.

The localized wavefunctions at the wells of the periodic potential
can be approximated by Wannier functions, i.e., the Fourier
transform of Bloch functions. Given the completeness of the
Wannier basis, any solution of the NLS equation with a periodic
potential can be expressed as
$u (x,t)=\sum_{n, \alpha }c_{n,\alpha }(t)w_{n,\alpha }(z)$,
 where
%the indices
$n$ and $\alpha$ label wells and bands of the periodic potential,
respectively.

Substituting the above expression into the dynamical equation,
and using the orthonormality of the Wannier basis as indicated
in \citeasnoun{wann1},
we obtain a set of differential equations for the coefficients.
Upon suitable decay of the Fourier coefficients and
the Wannier functions' prefactors (which can be systematically checked
for given potential parameters),
the model can be reduced to
\begin{eqnarray}
i {\frac{dc_{n, \alpha}}{dt}} =\hat{\omega}_{0,\alpha}c_{n,\alpha}
+ \hat{\omega}_{1,\alpha }\left(c_{n-1,\alpha} +c_{n+1,\alpha}\right)
%\nonumber \\
+ \sum_{\alpha_1, \alpha_2, \alpha_3}
 W^{nnnn}_{\alpha \alpha_1 \alpha_2 \alpha_3 }  c_{n,\alpha_1}^{\ast}
 c_{n,\alpha_2} c_{n,\alpha_3},
\label{chap01:i-ii}
\end{eqnarray}
where $W^{n n_1 n_2 n_3}_{\alpha \alpha_1 \alpha_2
\alpha_3} = \int_{-\infty}^{\infty}
w_{n,\alpha}w_{n_1,\alpha_1}w_{n_2,\alpha_2}w_{n_3,\alpha_3} dx$
is the Wannier function overlap integral and
\begin{equation}
\hat{\omega}_{n,\alpha }=\frac{L}{2\pi
}\int_{-\pi /L}^{\pi /L}E_{\alpha }(k)e^{-iknL}dk\, ,
\end{equation}
is the Fourier transform of the energy $E_{\alpha}(k)$ of
the Bloch mode corresponding to wavenumber $k$ for a potential
of spatial period $L$.
This degenerates into the so-called tight-binding model
\begin{eqnarray}
\hskip-1cm
i \frac{dc_{n,\alpha}}{dt} =\hat{\omega}_{0,\alpha}c_{n,\alpha}
+ \hat{\omega}_{1,\alpha}\left(c_{n-1,\alpha}
+c_{n+1,\alpha}\right)
%\nonumber \\
+ W^{nnnn}_{1111}  |c_{n,\alpha}|^2 c_{n,\alpha},
\label{chap01:TB}
\end{eqnarray}
i.e., the DNLS equation, when restricting consideration to the first band.
Higher-dimensional versions of the latter are of course physically
relevant models  and have, therefore, been used in various studies
concerning quasi-2D and 3D BECs confined
in strong optical lattices as illustrated in \citeasnoun{emergent}
and similarly in 2D waveguide
arrays as summarized in \citeasnoun{pgkbook}.

\begin{figure}[t]
\begin{center}
\includegraphics[width=4.5cm]{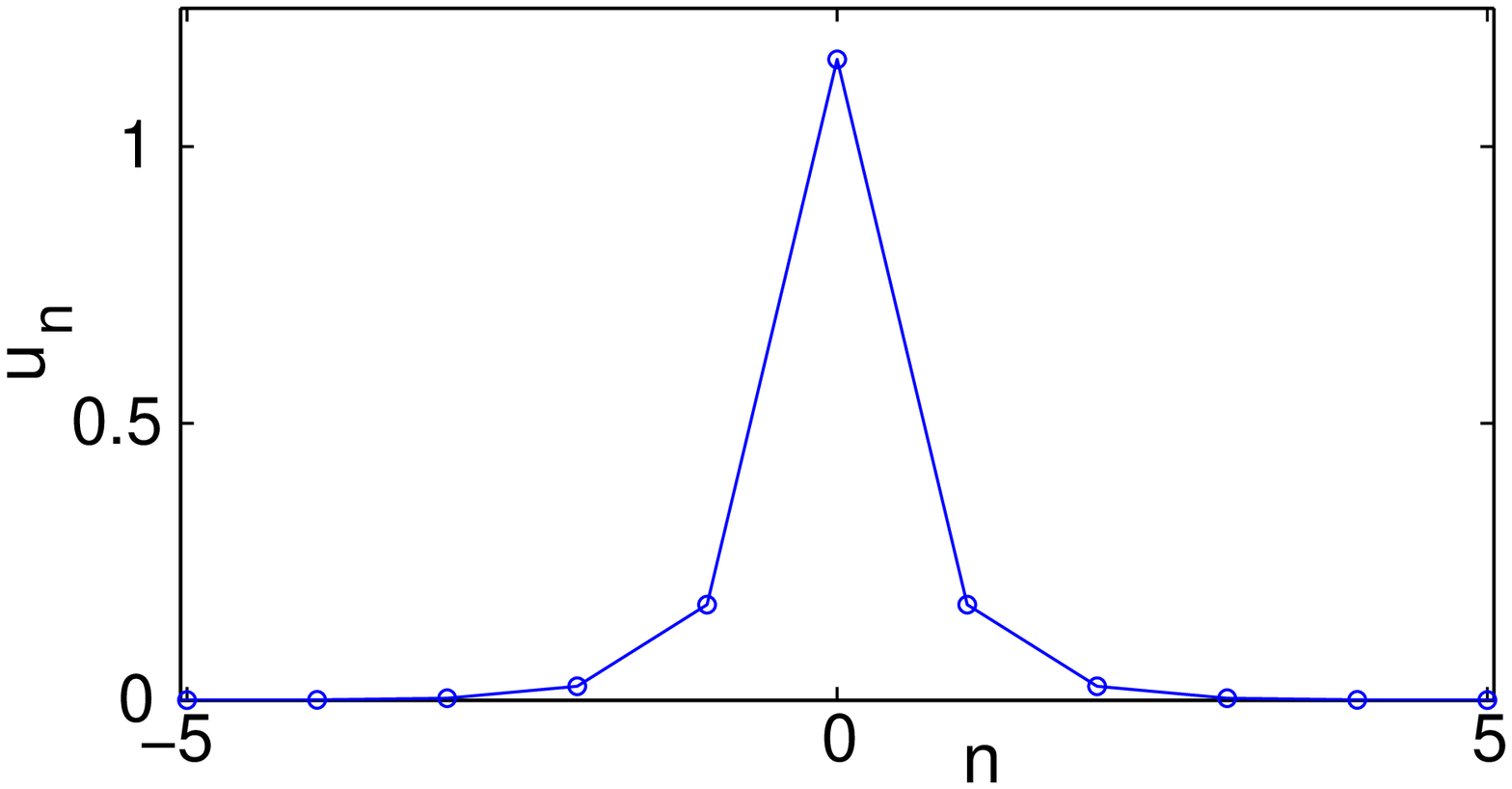}
\includegraphics[width=4.5cm]{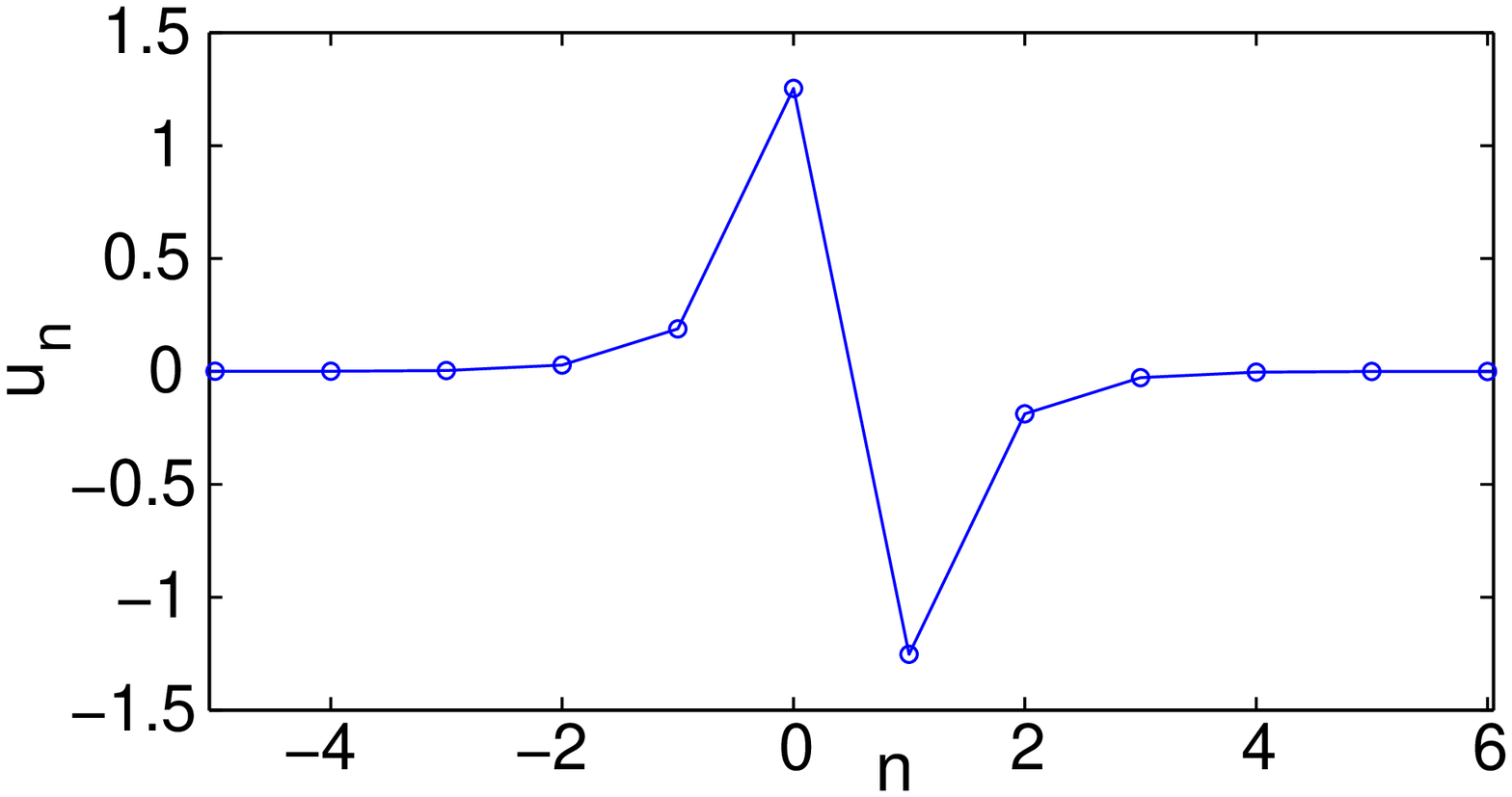}
\includegraphics[width=4.5cm]{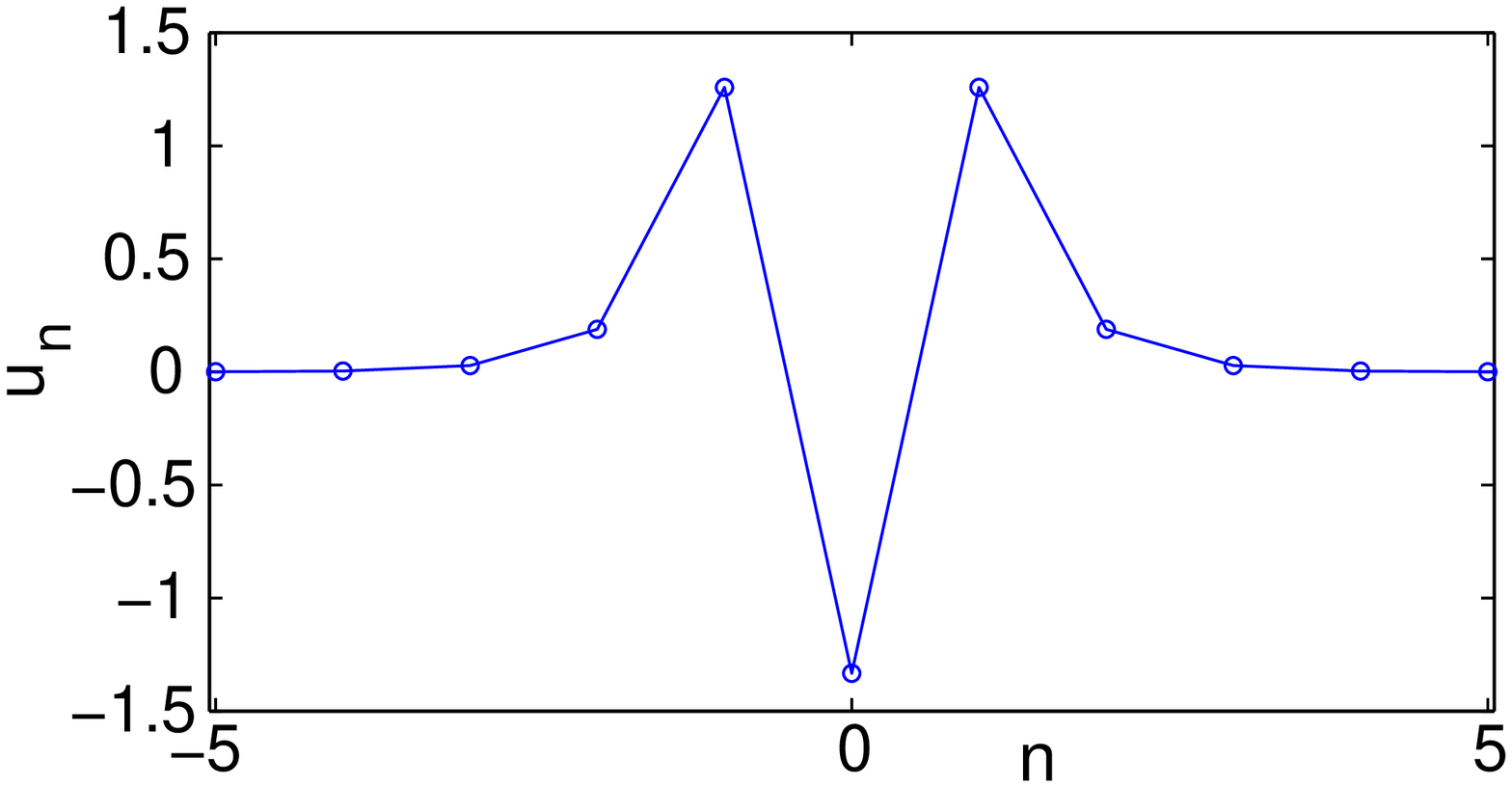}
\end{center}
%\begin{figure}[t]
\begin{center}
\includegraphics[width=4.5cm]{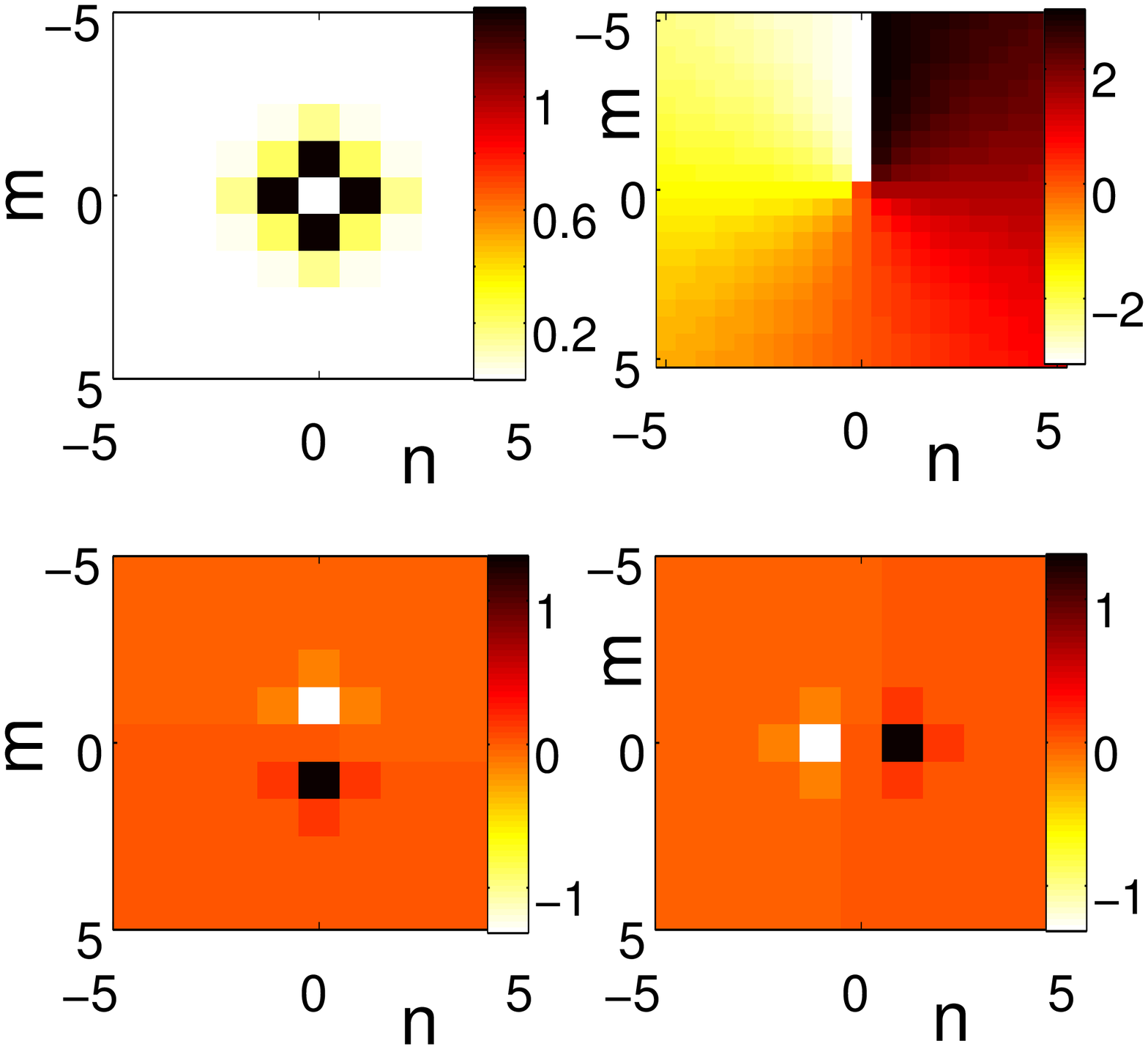}
\includegraphics[width=4.5cm]{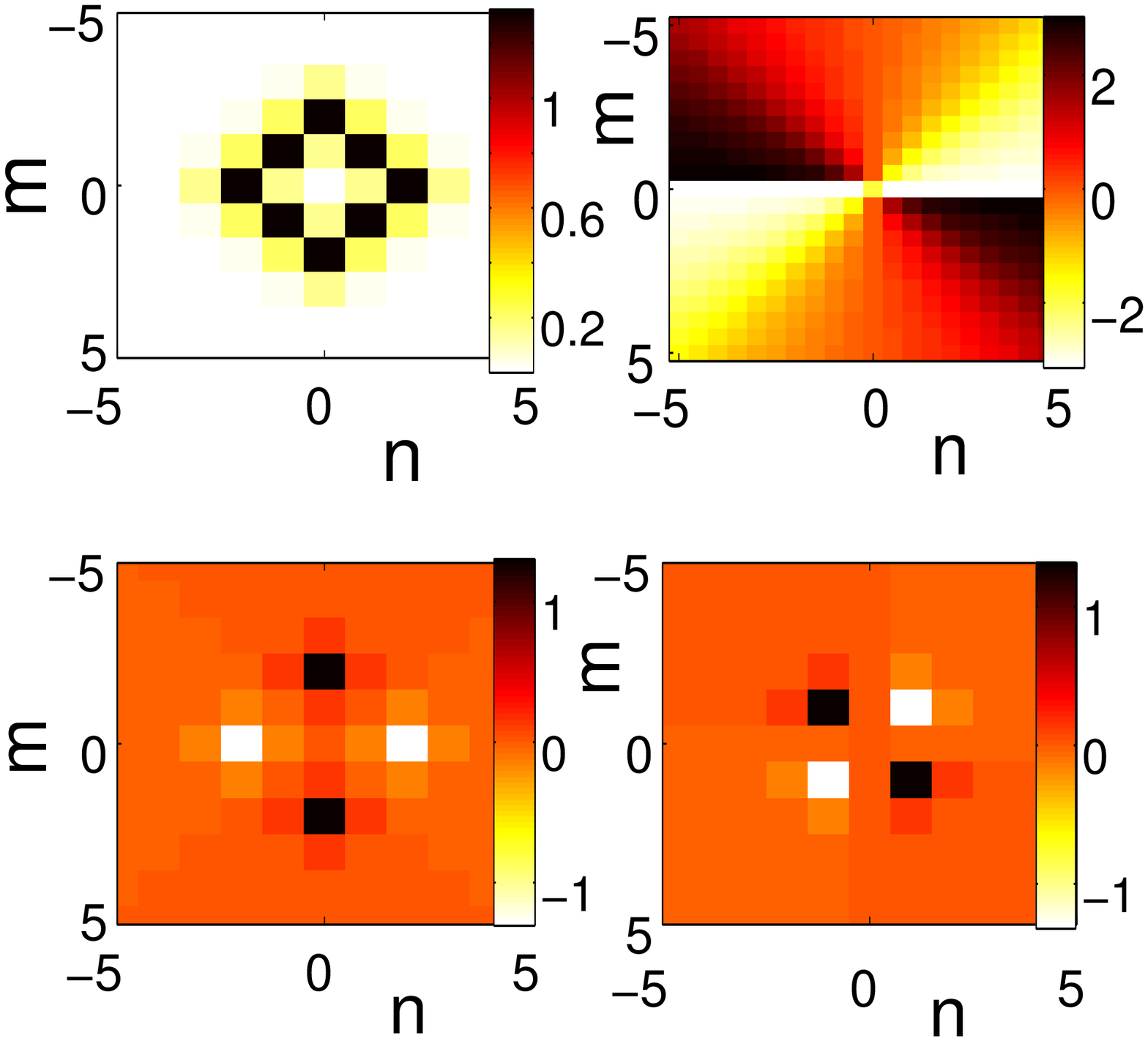}
\includegraphics[width=4.25cm]{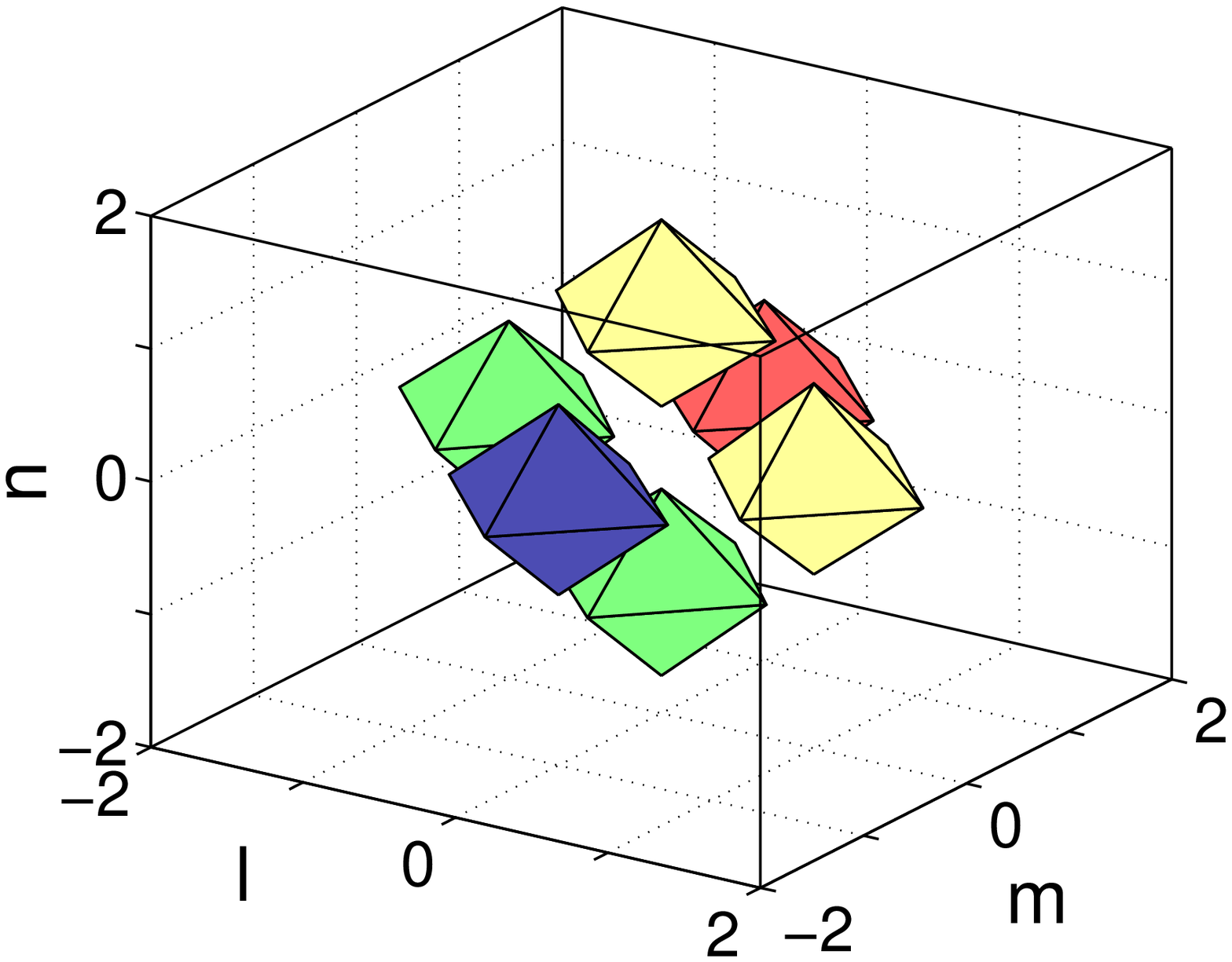}
\end{center}
%\vskip-0.4cm
\caption{(Color online) A sampler of solutions of the discrete nonlinear
Schr{\"o}dinger equation in 1-, 2- and 3-spatial lattice dimensions.
The top left shows the always stable fundamental single soliton.
The top middle and right show the out-of-phase two- and three-site
excitations (which can be stable near the anti-continuum limit).
The bottom left (4-panel set including the amplitude, phase, 
real and imaginary part contour plots of the solution) 
shows the prototypical vortex of topological
charge $S=1$ (which is also stable near the AC limit). The bottom
middle panel contains a
rhombic $S=2$ generalization thereof with $8$ sites (which may also
be stable), while the bottom right contains the also stable near
the AC limit 6-site diamond three-dimensional configuration.
}
%(Color online)
%A typical example of a two-site configuration is shown in the left
%panels of the figure, and a corresponding one of a three-site configuration
%in the right panels. The top panels show a solution (for a particular value
%of $\epsilon$) and its corresponding spectral plane $(\lambda_r,\lambda_i)$
%of eigenvalues $\lambda=\lambda_r+i\lambda_i$, while the bottom ones
%show the dependence of the relevant real eigenvalues as a function of
%the inter-site coupling $\epsilon$ obtained analytically (dashed/red lines)
%and numerically (solid/black lines).
%}
\label{rev_fig6}
\end{figure}

The relevance of the DNLS model in this wide array of application areas
prompted an intense activity in investigating the fundamental
solitary wave solutions of the model. We now focus on analyzing these
prototypical solutions and their stability in the context of
  Eq.~(\ref{s5_45}). In particular, for stationary solutions,
upon multiplying Eq.~(\ref{s5_46}) by $v_n^{\ast}$ and
subtracting the complex conjugate
of the resulting equation leads to:
\begin{equation}
v_n^{\ast} v_{n+1}-v_n v_{n+1}^{\ast}= {\rm const.}
~~\Rightarrow~~ 2 {\rm arg}(v_{n+1})= 2 {\rm arg}(v_n),
\label{s5_47}
\end{equation}
for localized  solutions (i.e., vanishing as $n \rightarrow \pm \infty$).
Using the scaling freedom of the equation to set $\mu=1$
allows us to infer that the only states that will
persist for finite $\epsilon$ will be the ones containing sequences with
combinations of $v_n=\pm 1$ and $v_n=0$.
%A systematic computational
%classification of the simplest ones among these sequences and of
%their bifurcations is provided in
The work of~\citeasnoun{konodisc} provided a systematic numerical
classification of the most fundamental of the resulting sequences
and their bifurcations. Some of the prototypical examples of sequences
involving one, two and three sites are illustrated in Fig.~\ref{rev_fig6}.
%Notice that we are tackling
%here the focusing case of $s=-1$, however, the defocusing case
%of $s=1$ can also be addressed based on the same considerations
While in our discussion below, we restrict our consideration to the
case of focusing nonlinearity, our results can be extended to the
defocusing case [with the opposite sign of the nonlinearity in
Eq.~(\ref{s5_45})] via the so-called staggering transformation
$w_n=(-1)^n u_n$. This converts the defocusing nonlinearity into a
focusing one, with an appropriate frequency rescaling which can
be trivially absorbed in a phase or gauge transformation.

We now turn to the examination of stability.
Using the linearization ansatz of Eq.~(\ref{stab_dnls}) and a simple
eigenvector rotation leads to the
equivalent symplectic formalism of the linear stability problem
$J {\cal L} w = \lambda w$, where
${\cal L}$ is a diagonal, self-adjoint matrix with the
operators ${\cal L}_+$ and ${\cal L}_-$
given below in its diagonal entries  and $J$ is the symplectic
matrix. ${\cal L}_+$ and ${\cal L}_-$ are defined by:
\begin{eqnarray}
(1-3 v_n^2) a_n - \epsilon (a_{n+1}+a_{n-1}) &=& {\cal L}_+ a_n = - \lambda b_n
\label{s5_48}
\\[1.0ex]
(1-v_n^2) b_n - \epsilon (b_{n+1}+b_{n-1}) &=& {\cal L}_- b_n = \lambda a_n.
\label{s5_49}
\end{eqnarray}
Using once again the AC limit, we assume
a $v_n$ with $N$ ``excited'' (i.e., $\neq 0$) sites; it can then
be straightforwardly inferred that for
$\epsilon=0$ these sites correspond to eigenvalues $\lambda_{+}=-2$
for
${\cal L}_+$ and to ones with $\lambda_{-}=0$ for ${\cal L}_-$.
In turn, these imply
the existence of $N$ eigenvalue pairs  with $\lambda^2=0$ for the
full problem.
These $N$ vanishing eigenvalue pairs are potential sources of instability,
since $N-1$ of those will become nonzero, upon departure from
the AC limit, given that there is
a single symmetry left for nonzero $\epsilon$, namely the U$(1)$ invariance
i.e., the invariance with respect to phase.
The key issue for stability purposes is to
identify the location of these $N-1$ small eigenvalue pairs.
One can manipulate Eqs.~(\ref{s5_48})--(\ref{s5_49}) into the form:
\begin{equation}
{\cal L}_- b_n =-\lambda^2  {\cal L}_+^{-1} b_n
~~\Rightarrow~~ \lambda^2= -
\frac{(b_n, {\cal L}_- b_n)}{(b_n, {\cal L}_+^{-1} b_n)}.
\label{s5_50}
\end{equation}
In the vicinity of the AC limit, the effect of ${\cal L}_+$ is a multiplicative
one (by $-2$). Hence:
\begin{equation}
\lim_{\epsilon \rightarrow 0}(b_n, {\cal L}_+^{-1} b_n)=-\frac{1}{2}
~~\Rightarrow~~ \lambda^2 = 2 \gamma =2 (b_n, {\cal L}_- b_n).
\label{s5_51}
\end{equation}
In light of this calculation, the full problem
becomes equivalent to the determination of the spectrum
of ${\cal L}_-$. A crucial fact in that regard is that $v_n$ is an
eigenfunction of ${\cal L}_-$ with $\lambda_{-}=0$. Then,
direct use of the Sturm comparison
theorem for difference operators [see e.g.~\citeasnoun{levy}]
leads us to conclude that if
the number of sign changes in the solution at the AC limit is $m$
(i.e., the number of times that adjacent to a $+1$ is a $-1$
and next to a $-1$ is a $+1$), then $n({\cal L}_-)=m$ and therefore
from Eq.~(\ref{s5_51}), the number of imaginary eigenvalues
pairs of $J {\cal L} $ is $m$. This, in turn, implies that
the number of real eigenvalue pairs
is consequently $(N-1)-m$. Another immediate conclusion is that
unless $m=N-1$, i.e., unless adjacent sites are
out-of-phase with respect to each other, the solution will be immediately
unstable upon departure from the AC limit. This analysis
originally presented in \citeasnoun{pelid1} is also consistent
with the general eigenvalue count of~\citeasnoun{kks}.
% since
%$n({\cal L})-n(D)=(N+m)-1=(N+m-1)+2 \times m + 2 \times 0=
%k_r + 2 k_i^{-} + 2 k_c$ (it is straightforward to show by the
%definition of the Krein signature \citeasnoun{pelid1} to show
%that these $m$ imaginary pairs have negative Krein signature).
It should be noted that these $m$ imaginary eigenvalue pairs
end up possessing, so-called, negative Krein signature
as analyzed in~\citeasnoun{pelid1}.
This implies that upon collision with other eigenvalue pairs, such
as the ones stemming from the continuous spectrum, they will lead
to instability through a so-called Hamiltonian-Hopf bifurcation
and a complex eigenvalue quartet. In this problem, the continuous
spectrum eigenmodes of eigenfrequency of wavenumber $k$ are
characterized by the dispersion relation
$\lambda= \pm i (1 + 4 \epsilon \sin^2(k/2))$, and hence correspond
to a band of width $4 \epsilon$, separated by a unit distance
from the origin along the imaginary axis
of the spectral plane $(\lambda_r,\lambda_i)$ of the eigenvalues
$\lambda=\lambda_r + i \lambda_i$.

Equation (\ref{s5_51}) can also be used in a quantitative fashion to identify
the relevant eigenvalues perturbatively for the full problem.
%by
%considering the perturbed (originally zero when $\epsilon=0$)
In particular, this amounts to considering the
eigenvalues of ${\cal L}_-$ emanating from $0$
through the perturbed eigenvalue problem:
\begin{equation}
{\cal L}_{-}^{(0)} b_n^{(1)} = \gamma_1 b_n^{(0)} -
{\cal L}_{-}^{(1)} b_n^{(0)},
\label{s5_52}
\end{equation}
where ${\cal L}_-={\cal L}_-^{(0)} + \epsilon {\cal L}_-^{(1)} + {\cal O}(\epsilon^2)$
and a similar expansion has been used for the eigenvector $b_n$.
For these eigenvalues, we know that
 $\lambda_{-}= \epsilon \gamma_1 + {\cal O}(\epsilon^2)$.
Projecting the above equation
to all the eigenvectors of zero eigenvalue of ${\cal L}_-^{(0)}$, one
can explicitly convert Eq.~(\ref{s5_52}) into an eigenvalue
problem of the form $M c= \gamma_1 c$, as was shown in \citeasnoun{pelid1}.
The
matrix $M$ has off-diagonal entries:
$M_{n,n+1}=M_{n+1,n}=-\cos(\theta_{n+1}-\theta_n)$ and
diagonal entries $M_{n,n}=\cos(\theta_{n-1}-\theta_n)+
\cos(\theta_{n+1}-\theta_n)$.
The subsequent computation of the leading order correction $\gamma_1$
(to the eigenvalues of ${\cal L}_-$)
%and subsequently from Eq.~(\ref{s5_51}) to
%evaluate the corresponding
allows us to calculate the perturbed eigenvalues of the full problem
$\lambda=\pm \sqrt{2 \epsilon \gamma_1}$.

Let us consider as a case example the
one-dimensional configurations with two-adjacent
sites with phases $\theta_1$ and $\theta_2$. Then, the matrix $M$ becomes:
\begin{eqnarray}
M=
\left( \begin{array}{cc}
\cos(\theta_1-\theta_2) & -\cos(\theta_1-\theta_2) \\
-\cos(\theta_1-\theta_2) & \cos(\theta_1-\theta_2) \end{array} \right),
\label{s5_53}
\end{eqnarray}
which leads to
$\lambda^2=0$ and $\lambda^2=4 \epsilon \cos(\theta_1-\theta_2)$.
Notice that, as expected,
for same phase excitations ($\theta_1=\theta_2$), the configuration
is unstable due to a real eigenvalue pair,
while the opposite is true if $\theta_1=\theta_2 \pm \pi$.
This type of analysis is possible e.g. for 3-site configurations with
phases $\theta_{1,2,3}$ (or for that matter, for arbitrary numbers of
excited sites). In the 3-site case,  one of the eigenvalues of
$M$ is again $0$ (this is true for any configuration due to the
U$(1)$ invariance), while the other two are given by:
\begin{eqnarray}
\gamma_{1} &=& \cos(\theta_2 - \theta_1) + \cos(\theta_3 - \theta_2)
\nonumber
\\
& \pm & \sqrt{\cos^2(\theta_2 - \theta_1) - \cos(\theta_2 - \theta_1)
\cos(\theta_3 - \theta_2) + \cos^2(\theta_3-\theta_2)}.
\label{s5_54}
\end{eqnarray}

%Some of the examples of the accuracy of these theoretical
%predictions for some typical two-site and three-site configurations
%(in particular, the in-phase ones, which should have one and
%two real eigenvalue pairs respectively) are shown in Fig.~\ref{rev_fig6}.

This formulation of the existence and stability problems
has been generalized to different settings, such as
higher dimensions in \citeasnoun{pelid2}, and \citeasnoun{pelid3} or
multi-component systems analyzed e.g. in
\citeasnoun{pelimulti}. Arguably, the principal
difference that arises in the higher dimensional settings is
that the wave profile may contain sites excited over a {\it contour}
(as opposed to along a straight line). In that case, for the
$N$ excited sites around the contour, the persistence (Lyapunov-Schmidt)
conditions can be obtained as a generalization of Eq.~(\ref{s5_45})
that reads [see e.g. \citeasnoun{pelid2} and \citeasnoun{sukhor}]:
\begin{eqnarray}
\sin(\theta_1-\theta_2)=\sin(\theta_2-\theta_3)=\dots=\sin(\theta_N-\theta_1).
\label{s5_55}
\end{eqnarray}
This highlights a key difference of these higher dimensional settings,
namely that not only ``solitary wave'' structures with phases
$\theta \in \{0,\pi\}$
are possible, but also both symmetric and asymmetric vortex families
presented in
\citeasnoun{pelid2} and \citeasnoun{sukhor} may, in principle, exist [a note
of caution, however, is that
Eq.~(\ref{s5_55}) provides only the leading order persistence condition and one
would need to also verify
the corresponding conditions to higher order to confirm
that such solutions persist as was discussed in \citeasnoun{pelid2}].
These vortex solutions had
been predicted numerically earlier in \citeasnoun{boris1} and \citeasnoun{boris2}
and have been
observed experimentally in the optical setting of photorefractive
crystals by \citeasnoun{vortex} and \citeasnoun{vortex1}. The stability
analysis is also possible for these higher dimensional structures, although
the relevant calculations are technically considerably more involved.
In fact, these types of calculations are possible not only for square
lattices, as in the cases highlighted above, but for more complex lattice
settings as well including hexagonal and honeycomb lattices as shown in
\citeasnoun{kody}. A number of interesting conclusions arise therein
including e.g. the instability of a lower topological charge (with
vorticity $S=1$) vortex and the stability of a higher topological
charge one (with $S=2$); such results have been recently also confirmed
experimentally by \citeasnoun{richter}. Some case examples of
interesting two-dimensional structures are given in Fig.~\ref{rev_fig6},
including a prototypical vortex cross of topological charge $S=1$,
as well as a potentially stable $S=2$ generalization thereof in a
rhombic configuration discussed recently in \citeasnoun{johs2} and
\citeasnoun{kodys2}.

However, it should also be noted that the relevant
theory can be formulated in an entirely general manner: we give the
outline below, as well as illustrate some prototypical higher-dimensional
(i.e., 3D) structures.
In the multi-dimensional case, the problem of existence of
stationary standing wave solutions can be
formulated through the vanishing of the vector field ${\bf F}_n$ of the form:
\begin{equation}
\label{nonlinear-vector-field} {\bf F}_n(\mbox{\boldmath
$\phi$},\epsilon) = \left[
\begin{array}{c}
(1 - |\phi_n|^2) \phi_n - \epsilon \Sigma \phi_n \\ [1.0ex]
(1 - |\phi_n|^2) {\phi}_n^{\ast} - \epsilon \Sigma
{\phi}_n^{\ast}  \end{array} \right].
\end{equation}
Upon defining the matrix operator:
\begin{eqnarray}
\label{energy} {\cal H}_{n} &=&
\left( \begin{array}{cc}
1 - 2|\phi_n|^2 & - \phi_n^2 \\ [1.0ex]
- {\phi_n^{\ast}}^2 & 1 - 2 |\phi_n|^2
\end{array} \right)
\nonumber
\\
&-& \epsilon \left( s_{+e_1} + s_{-e_1} +
s_{+e_2} + s_{-e_2} + s_{+e_3} + s_{-e_3} \right) \left(
\begin{array}{cc} 1 & 0 \\ 0 & 1 \end{array} \right),
\end{eqnarray}
where the $s_{\pm e_i}$ denote the shift operators along the respective
directions and $e_i$ are the corresponding unit vectors in each
of the three lattice directions, the stability problem reads $\sigma {\cal H}
\mbox{\boldmath $\psi$} = i
\lambda \mbox{\boldmath $\psi$}$. Here each $2 \times 2$ block of the
matrix $\sigma$ is
the diagonal matrix with elements $(1,-1)$ along the diagonal for 
each node $n$; see \citeasnoun{pelid3} for a detailed discussion.
The existence problem, on the other hand,
is connected to ${\cal H}$ through:
${\cal H} = D_{\mbox{\boldmath $\phi$}} {\bf
F}(\mbox{\boldmath $\phi$},0)$. At the AC limit of $\epsilon=0$
$$
({\cal H}^{(0)})_n = \left[ \begin{array}{cc} 1 & 0 \\ 0 & 1
\end{array} \right], \; n \in S^{\perp}, \qquad
({\cal H}^{(0)})_n = \left[ \begin{array}{cc} -1 & -e^{2 i \theta_n} \\
-e^{-2 i \theta_n} & -1 \end{array} \right], \; n \in S,
$$
where $S$ is the set of excited sites. Then the eigenvectors
of zero eigenvalue will be of the form:
$$
({\bf e}_n)_k = i \left[ \begin{array}{c} e^{i \theta_n} \\ -
e^{-i \theta_n}\end{array} \right] \; \delta_{k,n}.
$$
It is helpful to define the projection operator:
\begin{equation}
({\cal P} {\bf f})_n = \frac{\left( {\bf e}_n,
{\bf f} \right)}{\left( {\bf e}_n, {\bf e}_n \right)} = \frac{1}{2i}
\left( e^{-i \theta_n} ({\bf f}_1)_n - e^{i \theta_n}({\bf f}_2)_n
\right), \qquad n \in S,
\end{equation}
and to decompose the solution as:
\begin{eqnarray}
\mbox{\boldmath $\phi$} = \mbox{\boldmath
$\phi$}^{(0)}(\mbox{\boldmath $\theta$})
%+ \sum_{n \in S} \alpha_n
%{\bf e}_n
+ \mbox{\boldmath $\varphi$} \in X.
\end{eqnarray}
Then,
one can obtain the Lyapunov-Schmidt persistence conditions
analyzed in \citeasnoun{pelid3} as:
\begin{equation}
\label{g-definition} {\bf g}(\mbox{\boldmath $\theta$},\epsilon) =
{\cal P} {\bf F}(\mbox{\boldmath $\phi$}^{(0)}(\mbox{\boldmath
$\theta$})+\mbox{\boldmath $\varphi$}(\mbox{\boldmath
$\theta$},\epsilon),\epsilon) = 0.
\end{equation}
%This leads to the persistence theorem \citeasnoun{pelid3}:
As proved in \citeasnoun{pelid3}, the configuration $\mbox{\boldmath
$\phi$}^{(0)}(\mbox{\boldmath $\theta$})$
can then be continued to the domain $\epsilon \in {\cal O}(0)$
(i.e., for couplings in the neighborhood of the AC limit)
if and
only if there exists a root $\mbox{\boldmath $\theta$}_*$
% \in {\cal
%T}$
of the vector field ${\bf g}(\mbox{\boldmath
$\theta$},\epsilon)$. Moreover, if the
root $\mbox{\boldmath $\theta$}_*$ is analytic in $\epsilon \in
{\cal O}(0)$ and $\mbox{\boldmath $\theta$}_* = \mbox{\boldmath
$\theta$}_0 + {\cal O}(\epsilon)$, the solution $\mbox{\boldmath
$\phi$}$ of the difference equation
%(\ref{3difference})
is
analytic in $\epsilon \in {\cal O}(0)$, such that
\begin{equation}
\label{Taylor-series-phi} \mbox{\boldmath $\phi$} = \mbox{\boldmath
$\phi$}^{(0)}(\mbox{\boldmath $\theta$}_*) + \mbox{\boldmath
$\varphi$}(\mbox{\boldmath $\theta$}_*,\epsilon) = \mbox{\boldmath
$\phi$}^{(0)}(\mbox{\boldmath $\theta$}_0) + \sum_{k=1}^{\infty}
\epsilon^k \mbox{\boldmath $\phi$}^{(k)}(\mbox{\boldmath
$\theta$}_0).
\end{equation}
Within the same formulation, one can establish a general stability theory,
provided that the relevant solution persists for $\epsilon \neq 0$.
%More specifically, let the solution of interest persist for $\epsilon \neq 0$.
If the
operator ${\cal H}$ has a small eigenvalue $\mu$ of multiplicity $d$,
such that $\mu = \epsilon^k \mu_k + {\cal O}(\epsilon^{k+1})$, then
the full Hamiltonian
eigenvalue problem admits $(2d)$ small
eigenvalues $\lambda$. These are such that $\lambda = \epsilon^{k/2}
\lambda_{k/2} + {\cal O}(\epsilon^{k/2+1})$, where non-zero values
$\lambda_{k/2}$ are found from
\begin{eqnarray}
\label{reduced-problem-1} && \mbox{odd $k$:} \quad {\cal M}^{(k)}
\mbox{\boldmath $\alpha$} = \frac{1}{2} \lambda_{k/2}^2 \mbox{\boldmath $\alpha$}, \\[1.0ex]
\label{reduced-problem-2} && \mbox{even $k$:} \quad {\cal M}^{(k)}
\mbox{\boldmath $\alpha$} + \frac{1}{2} \lambda_{k/2} {\cal
L}^{(k)} \mbox{\boldmath $\alpha$} = \frac{1}{2} \lambda_{k/2}^2
\mbox{\boldmath $\alpha$},
\end{eqnarray}
where
\begin{eqnarray}
{\cal M}^{(k)} &=&~ D_{\mbox{\boldmath $\theta$}} {\bf g}^{(k)}(\mbox{\boldmath $\theta$}_0),
\nonumber
\\[1.0ex]
~{\cal L}^{(k)} &=&~ {\cal P} \left[ {\cal H}^{(1)}
\mbox{\boldmath $\Phi$}^{(k')}(\mbox{\boldmath $\theta$}_0) + ...
+ {\cal H}^{(k'+1)} \mbox{\boldmath $\Phi$}^{(0)}(\mbox{\boldmath
$\theta$}_0) \right],
\nonumber
\end{eqnarray}
and $k'=(k-1)/2$. For more details, we refer the interested
reader to~\citeasnoun{pelid3}. A prototypical example
of a configuration (the so-called discrete diamond)
that the above theory predicts
to exist and be dynamically robust near the AC limit in 3D
is illustrated in the bottom right panel of Fig.~\ref{rev_fig6}.
Other such configurations can be found e.g. in \citeasnoun{pelid3},
as well as in \citeasnoun{ricardo3d1} and \citeasnoun{ricardo3d2}.

%In Fig. we show typical examples of 2D and
%3D configurations that satisfy the
%persistence conditions formulated above. The former configuration is a vortex
%of topological charge $S=2$ (i.e., its phase changes uniformly by
%$\pi/2$ from each site to the next so that it changes from $0$
%to $4 \pi$ around the discrete contour of the solution). This structure is
%unstable due to a real eigenvalue pair that is theoretically predicted
%from Eqs.~(\ref{reduced-problem-1})--(\ref{reduced-problem-2}) to be
%$\lambda=\pm \sqrt{\sqrt{80}-8} \epsilon$ (while it also has a
%pair of simple eigenvalues $\lambda \pm i \epsilon \sqrt{\sqrt{80}+8}$,
%a quadruple eigenvalue $\lambda \pm i \epsilon \sqrt{2}$ and a single
%eigenvalue of higher order). The latter configuration is a three-dimensional
%diamond configuration
%(a quadrupole in the plane with phases $0, \pi, 0, \pi$ and two
%out-of-plane sites with phases $\pi/2$ and $3 \pi/2$). This is a stable
%3D structure with a single eigenvalue $\lambda= \pm 4 i \epsilon$,
%a triple eigenvalue $\lambda=\pm 2 i \epsilon$ and an eigenvalue of
%higher order according to
%Eqs.~(\ref{reduced-problem-1})--(\ref{reduced-problem-2}). Notice in
%both cases the remarkable agreement between the theoretical prediction
%for the eigenvalues (dashed lines) and the full numerical results (solid
%lines).

\subsection{Future Perspectives}
\label{future}

Although, based on these recent results, much  is presently
understood about the statics and stability of DNLS type systems,
a number of themes still remain relatively unexplored. We
 now present a few of the themes that we believe hold significant
interest for future studies.

\subsubsection{\it A general theory and applications of
long-range interactions.}

While the setting of local (i.e., nearest-neighbor) interactions
has been substantially explored in the DNLS case, this can far less
be argued to be the case for the setting of non-nearest neighbor
effects. Some early investigations of one dimensional
settings identified the role of non-nearest neighbor
interactions in creating bistability of the fundamental
soliton solutions and leading to interesting possibilities
for switching among stable branches (upon suitable perturbations)
as in \citeasnoun{johold1} and \citeasnoun{johold2}. However, a more
systematic investigation of existence and stability issues,
especially for more complicated solutions was not offered.

More recently, a number of physically-minded works have argued
the relevance of the inclusion of such non-nearest neighbor
interactions in a variety of settings, involving predominantly
optical waveguide arrays. For instance, the work of \citeasnoun{efremdnc}
argued that a zigzag waveguide array could be considered as a
quasi-one-dimensional chain in which the relative strength of nearest-neighbor
and next-nearest-neighbor interactions can be tuned on the basis of the
relevant geometry. Corresponding ideas for involving non-nearest-neighbor
interactions have also been generalized to two-dimensional settings.
Although limited to the linear propagation at least in the realm
of \citeasnoun{pertsch}, even in that setting,
there is interesting phenomenology
arising from the competition of the different types of interactions,
leading potentially to diffraction-free propagation for suitable
wavenumbers in the center of the Brillouin zone. On the nonlinear
side, the work of \citeasnoun{pgkpla} indicates that both the
existence problem and the stability may be crucially modified with
respect to the standard DNLS case by the presence of nonlocality.
A typical example of the former involves the emergence of 1D
solutions that have arbitrary (i.e., different than $0, \pi$) phases.
A typical example of the latter involves the stabilization  of
unstable structures [such as
a prototypical class of $S=2$ vortices in a square lattice
of \citeasnoun{pelid2}]. It should also be noted that another
area that may enhance the relevance of long-range interaction
considerations is that of dipolar Bose-Einstein condensates (such as $^{52}$Cr)
in the presence of optical lattices in atomic physics; see e.g. the
recent review of \citeasnoun{lahaye}.

These directions and findings seem to warrant a more systematic 
investigation of
the effects of nonlocality, ideally as a general function of properties
(e.g. the ``interaction range'') of the kernel. Such a general setting
could be:
\begin{eqnarray}
i \dot{u}_n=-\epsilon \sum_{m=1}^N a_{nm} u_m - |u_n|^2 u_n
\label{lri}
\end{eqnarray}
for different types of interaction kernels $a_{nm}$ (although
the nonlocality could also in principle, or additionally
be imposed on the nonlinear term).
It would appear to
be timely and relevant to explore
how existence, stability and dynamics of solitary waves
are progressively modified, as we depart from the well-understood
local limit.

\subsubsection{\it Intersite Lattices, Discretizations, Symmetries
and Traveling Waves.}

One of the directions that also hold promise in the setting
of DNLS models is that of different types of discretizations
and their properties in comparison to the more fundamentally
well-understood model with the centered-difference Laplacian
and the onsite cubic nonlinearity. As a motivating example
in this direction, we briefly discuss the recent report
of \citeasnoun{lafortus} which examined the two-dimensional
Ablowitz-Ladik model of the form:
\begin{eqnarray}
i\dot u_{n,m}  =  &-& \varepsilon \left( {u_{n + 1,m}  + u_{n - 1,m}  + u_{n,m + 1}  + u_{n,m - 1}  - 4u_{n,m} } \right)
\nonumber
\\
&+& \frac{\sigma }{4}\left| u \right|^2 \left( {u_{n + 1,m}  + u_{n - 1,m}  + u_{n,m + 1}  + u_{n,m - 1} } \right).
\label{ALEqn}
\end{eqnarray}
The fundamental difference of this, so-called AL-NLS, model from the standard
DNLS is that
in addition to the centered-difference approximation of the
Laplacian,
a nearest-neighbor average is used to discretize
the cubic nonlinearity of the continuum limit $ \sigma |u|^2 u$
(instead of a local term $\sigma |u_{n,m}|^2 u_{n,m}$ in the DNLS).
In the one-dimensional setting, such a centered-difference
discretization of the nonlinearity is completely integrable, as shown by
\citeasnoun{AL76}, giving rise to exact hyperbolic-secant
solitonic solutions that can travel at arbitrary speeds.
This well-known (since the 1970s) fact already showcases the special
properties that can emerge upon different types of discretization.

However,
the recent work of \citeasnoun{lafortus} illustrated that surprising
features may arise because of such discretizations in higher-dimensional
settings as well. In particular, in the 2D case, it was found
that the DNLS solitons become unstable at some critical
threshold (e.g. of the coupling strength $\epsilon$) as
the continuum limit
is approached (e.g. by increasing $\epsilon$). The solitons
then remain unstable all the way to the continuum limit
(of $\epsilon \rightarrow \infty$)
with the relevant instability eigenvalue approaching the origin
of the spectral plane. This is because the 2D continuum NLS model
is critical i.e., it is marginally unstable with respect to collapse.
In the critical case, the continuum model is invariant under rescaling,
see e.g. the relevant analysis of
\citeasnoun{sulem}. On the other hand, the AL-NLS approaches
this continuum limit in a completely different way;
in particular, while the AL-NLS two-dimensional solitons
become unstable within a narrow range of parameters (such
as $\epsilon$), they
subsequently become restabilized and remain stable as the
continuum limit is approached. This suggests the remarkable fact that while the
two models (DNLS and AL-NLS) are identical at the limit,
infinitesimally close to the limit, their dynamics is
substantially different. While the DNLS solitons are exponentially
(yet weakly) unstable, the AL-NLS ones are dynamically stable.

Further consideration of this feature indicates that it is a
particular trait of critical settings, which are
at the very special
separatrix between subcritical settings where the solitary waves
are dynamically stable and supercritical ones, where the waves
are exponentially unstable. In this critical case, the linear
spectrum possesses an additional zero
eigenvalue pair (associated with the pseudo-conformal invariance),
which permits the reshaping of the solution under
the action of the group of rescalings, and hence paves the way
for the emergence of self-similar collapse. Discreteness
can then shift this pair along the imaginary axis or along
the real axis.
The AL-NLS discretization turns out to be a prototypical example
whereby the eigenvalue pair formerly associated with the pseudo-conformal
invariance is perturbed in a stable way (moves along the imaginary
axis of the spectral plane), upon discretization and hence this
model allows infinitesimally small spacings to give rise to
{\it collapse-free} dynamics.

The above discussion of issues pertaining to symmetry raises
the more general question of devising different types of
discretizations and examining their symmetry properties and
the connection of these to the dynamical features of solitary
waves. Assuming that discretizations of the continuum NLS
model will respect the phase/gauge-invariance of that model,
another key symmetry whose impact on discretizations has been
examined fairly extensively recently is the invariance with
respect to translations. Motivated by the early work of
\citeasnoun{speight1}, \citeasnoun{speight2} and that of
\citeasnoun{kevrekidis03}, there has been a considerable
volume of literature developed recently on the subject of
suitable discretizations of Klein-Gordon, as well as of
NLS models which preserve (for their stationary solutions)
the property of translational invariance; see e.g., \citeasnoun{peli06}
and references therein. It is well-known that such an invariance
is absent in the standard DNLS model which admits single-humped
discrete solitary waves which can only be centered on a site
or half-way through between two sites (but not other such types
of waveforms, contrary to what is the case with the translationally
invariant models where the solitary waves can be centered anywhere).
In this context, and for the NLS-model discretizations the
work of \citeasnoun{peli06} unified some of the earlier findings
and methods, illustrating that the most general translationally
invariant discretization of the cubic model is of the form:
\begin{eqnarray}
i\dot u_{n}  &\!\!=\!\!&
-\varepsilon \left( {u_{n + 1}  + u_{n - 1}  - 2 u_{n} } \right) - (1-\chi-4 \xi - 2 \eta) |u_n|^2 (u_{n+1} + u_{n-1})
- \chi u_n^2 ({u}^{\star}_{n+1} + {u}^{\star}_{n-1})
\nonumber
\\[1.0ex]
&&- \xi \left[(2 |u_n|^2 + |u_{n+1}|^2 + |u_{n-1}|^2) u_n
+ (u_{n+1}^{\star} u_{n-1} + u_{n+1} u^{\star}_{n-1}) u_n
+ (u_{n+1}^2 + u_{n-1}^2) u_n^{\star} \right]
\label{ALEqn2}
\\[1.0ex]
&&-\eta (|u_{n+1}|^2 + |u_{n-1}|^2) (u_{n+1} + u_{n-1})
- \nu \left[u_{n+1}^2 u^{\star}_{n-1} +  u_{n-1}^2 u^{\star}_{n+1}
\nonumber
-  |u_{n+1}|^2 u_{n-1} - |u_{n-1}|^2 u_{n+1} \right],
\end{eqnarray}
where $(\chi,\xi,\eta,\nu)$ are real-valued parameters. When all
four parameters vanish, one recovers the AL-NLS limit.

This discussion, in turn, leads to the interesting additional
question of the possibility of identifying traveling wave
solutions in DNLS (and related) lattices. This question was
considerably controversial during the previous decade with
many contradictory results and conjectures; see e.g. the
discussion in \citeasnoun{melvin06} and references therein.
Starting from the work of \citeasnoun{gomez04}, which
illustrated numerically that in DNLS such traveling waves cannot be
localized, this question was subsequently answered both
in the DNLS setting and in that of more complex models
such as the saturable nonlinear proposed in \citeasnoun{ljupco} in the
form:
\begin{eqnarray}
i \dot{u}_n = - \epsilon \Delta_2 u_n + \frac{1}{1 + |u_n|^2} u_n.
\label{sat}
\end{eqnarray}
What the direct numerical simulations of \citeasnoun{ljupco}
initially suggested and which was corroborated numerically
in the work of \citeasnoun{melvin06}, \citeasnoun{melvin07}
and quasi-analytically through the asymptotic expansions
of \citeasnoun{barashenkov07} was that indeed in DNLS,
the resonance of such traveling wave (``embedded soliton'')
type solutions with the (modified, for traveling solutions)
continuous spectrum gave
rise to ``nanoptera'' i.e., solutions with a non-vanishing
tail. However, in different types of nonlinearities such as
the saturable or cubic-quintic ones, these works showed that
it is possible to make the prefactor of such a resonance
(the so-called Stokes constant) vanish, thereby producing
non-generic, yet exact, exponentially localized traveling
solutions in these models.

The above findings, in turn, beg the somewhat
open-ended question: is there some simple diagnostic which may
indicate the existence of such traveling solutions in a model?
The works of \citeasnoun{ljupco} and \citeasnoun{melvin06}
[see also the studies of \citeasnoun{chrish1} and
\citeasnoun{chrish2} for corresponding discussions of the
cubic-quintic discrete model]
hinted at the use of the so-called Peierls-Nabarro barrier,
i.e., the energy difference between on-site and inter-site
centered solutions. In particular, they argued that vanishings
of this barrier should be physically expected to be connected
with the possibility of such traveling solutions. However,
this connection has not been made rigorous, as of yet.
More generally, exploring systematically the connection
of Peierls-Nabarro barriers, the existence of traveling solutions,
the calculation of the Stokes constant, and the potential for
underlying symmetries (including a possible semi-discrete type
of Galilean invariance) could be extremely interesting topics
for further exploration over the next decade. Understanding such
properties either on a model-specific basis, or, ideally, based
on more general/fundamental principles is an important open
direction for these nonlinear dynamical lattices.
At this point, we should also mention in passing the very
interesting recent work of \citeasnoun{barashenkov08}. There,
the examination of exceptional discretizations [bearing an
effective translational invariance through the map type approach
of \citeasnoun{peli06} and \citeasnoun{barashenkovpeli}] led to some
ingenious suggestions on how to discretize so as to preserve
genuinely traveling localized excitations i.e., kinks in the
discrete sine-Gordon and discrete $\phi^4$ settings. It would
be extremely worthwhile for such methodologies to be brought to
bear also in DNLS and related settings.

\subsubsection{\it Statistical Mechanics of the DNLS and Approach to
Equilibrium.}

One of the questions that has received a somewhat limited amount
of attention is that of the statistical mechanics of the DNLS model
and how a given initial condition approaches its asymptotic (equilibrium
or near-equilibrium) dynamics. One of the early studies of such
questions for the DNLS lattice appeared in \citeasnoun{sec15.RCKGJ}.
There,
a regime in phase
space was identified
wherein regular statistical mechanics considerations apply, and
hence, thermalization was observed numerically and explored analytically
using regular, grand-canonical, Gibbsian equilibrium measures.
However, the nonlinear dynamics of the problem renders permissible the
realization of regimes of phase space which would formally correspond to
``negative temperatures'' in the sense of statistical mechanics.
The novel feature of these states was found to be that
the energy spontaneously localizes in certain lattice sites
forming breather-like excitations (as observed numerically and
experimentally).
Returning to statistical mechanics, such
realizations are not
possible [since the Hamiltonian is unbounded, as is seen by
a simple scaling argument similar to the continuum case studied in
\citeasnoun{sec15.LEB}] unless  the grand-canonical Gibbsian measure is
refined to
correct for the unboundedness. This correction
was argued in \citeasnoun{sec15.RCKGJ} to produce a discontinuity in the
partition function
signaling a phase transition which was identified numerically by the
appearance of discrete breathers.

More recently, the statistical understanding of the formation of
localized states and of the asymptotic dynamics of the DNLS equation
has been examined in the works of \citeasnoun{sec15.rumpf}
and \citeasnoun{sec15.johras}.
Using small-amplitude initial conditions, \citeasnoun{sec15.rumpf}
argued that the phase space of the system can be divided roughly
into two weakly interacting domains, one corresponding to the
low-amplitude fluctuations (linear or phonon modes), while the other
consists of the large-amplitude, localized mode nonlinear excitations.
Then, based on a simple
partition of the energy $H=H_{-}+ H_{+}$ and of the norm
$P=P_{-} + P_{+}$, into these two broadly (and also somewhat loosely)
defined fractions, one smaller than a critical
threshold (denoted by ``$-$'') and one larger than a critical threshold
(denoted by ``$+$''), it is possible to compute thermodynamic quantities
such as the entropy in this localization regime.
In particular, one of the key results of the work of
\citeasnoun{sec15.rumpf} is that,
for a partition of $K$ sites with large amplitude excitations
and $N-K$ sites with small amplitude ones, an expression is derived
for the total entropy (upon computing $S_-$, $S_+$ and a permutation
entropy due to the different potential location of the $K$ and $N-K$ sites).
This expression reads:
\begin{eqnarray}
S=N {\rm ln} \Omega + \frac{P_+^2}{2 E_+} {\rm ln} \Gamma,
\label{sec15.eq4}
\end{eqnarray}
where $\Omega=(4 P_-^2-E_-^2)/(4 A_- (N-K))$
and $\Gamma=2 P_- N/P_+ E_-$, while $K=P_+^2/(2 E_+)$.
While some somewhat artificial assumptions are needed to
arrive at the result of Eq.~(\ref{sec15.eq4}) (such as the
existence of a cutoff amplitude radius $R$ in phase space),
nevertheless, the result provides a transparent physical
understanding of the localization process. The contributions
to the entropy stem from the fluctuations
[first term in Eq.~(\ref{sec15.eq4})] and from the high amplitude peaks
(second term in the equation). However, typically the contribution
of the latter in the entropy is negligible, while they can
absorb high amounts of energy. The underlying premise is that
the system seeks to maximize its entropy by allocating the ideal amount of
energy $H_-$ to the fluctuations. Starting from an initial energy
$H_-$, this energy is decreased in favor of
localized peaks (which contribute very little to the entropy).
The entropy would then be maximized if eventually a single
peak was formed, absorbing a very large fraction of the energy while
consuming very few particles. Nevertheless, practically, this regime
is not reached ``experimentally'' (i.e., in the simulations).
This is because of the inherent discreteness of the system which leads
to a pinning effect of large amplitude excitations which cannot move
(and, hence, cannot eventually
merge into a single one) within the lattice. Secondly,
the growth of the individual
peaks, as argued in \citeasnoun{sec15.rumpf}, stops when the entropy gain due
to energy transfer to the peaks is balanced by the entropy loss due
to transfer of power. While placing the considerations of
\citeasnoun{sec15.rumpf}
in a more rigorous setting is a task that remains open for future
considerations, this conceptual framework offers considerable potential
for understanding the (in this case argued to be infinite, rather than
negative, temperature) thermal equilibrium state of coexisting
large-amplitude localized excitations and small-amplitude background
fluctuations.

On the other hand, the work of \citeasnoun{sec15.johras} extended
the considerations of the earlier work of \citeasnoun{sec15.RCKGJ}
to the generalized DNLS model of the form:
\begin{eqnarray}
i\dot u_n+ (u_{n+1}+u_{n-1})+ |u_n|^{2 \sigma} u_n=0,
\label{sec15.req201}
\end{eqnarray}
with $\sigma$ being a free parameter within the nonlinearity exponent.
Furthermore, in the work of \citeasnoun{sec15.johras}, the connection
of these DNLS considerations with the generally more complicated
Klein-Gordon (KG)
models was discussed. Much of the above mentioned phenomenology,
as argued in \citeasnoun{sec15.rumpf}, is critically particular to NLS
type models, due to the presence of the second conserved quantity,
namely of the $l^2$ norm; this feature is absent in the KG
lattices, where typically only the Hamiltonian is conserved.
\citeasnoun{sec15.johras} formalize the connection
of DNLS with the KG lattices, by using the approximation of the latter via
the former through a Fourier expansion whose coefficients satisfy
the DNLS up to controllable corrections. Within this approximation,
they connect the conserved quantity of the KG model to the ones
of the DNLS model approximately reconstructing the relevant transition
(to formation of localized states) criterion discussed above.
However, in the KG setting this only provides a guideline for the
discrete breather formation process, as the conservation of the norm is no
longer a true but merely an approximate conservation law.
This is observed in the dynamical simulations of \citeasnoun{sec15.johras},
where although
as the amplitude remains small throughout the lattice the process
is well described by the DNLS formulation, when the 
discrete breathers of the
KG problem grow, they violate the validity of the DNLS approximation
and of the norm conservation; thus, a description of the asymptotic
state and of the thermodynamics of such lattices requires further
elucidation that necessitates a different approach. This is another
interesting and important problem for future studies.

It is also worth pointing out that many of the considerations
of works such as that of \citeasnoun{sec15.RCKGJ} or of
\citeasnoun{sec15.johras} are intrinsically one-dimensional in nature
and can not be straightforwardly generalized (per the nature
of the transfer integral technique used) to higher dimensions. Hence,
the issue of the asymptotic dynamics is perhaps more pressing
in such higher-dimensional settings, especially given their
connections to collapse (sufficiently close to the continuum limit).
But even in one-dimension,
numerous questions remain. One such interesting question was 
raised by the recent work of \citeasnoun{drossinos08}.
This work offers an understanding of the fundamental threshold
(for initial data supported on a single site) between discrete dispersion
for sub-critical initial amplitude
and nonlinearity-induced localization, for sufficiently high initial
amplitude. However, it also poses the question: once we
know localization will ensue, which ``member'' of the mono-parametric
family of localized solutions will the dynamics ``select'' for
the asymptotic relaxation state? Numerical experiments illustrate
that the initial condition will shed both some mass and some energy,
eventually converging to one particular member of that family of
stationary solutions. But what is missing is the guiding principle
of such a selection. Interestingly, while the single-site supported
initial data can give localization in the DNLS model (when it possesses
a sufficiently large amplitude), in the AL-NLS model, 
such data always leads to dispersion.
In the latter case, the machinery of the theory of integrability
can be brought to bear
to appreciate the effect of different types of compactly supported
initial conditions as shown in \citeasnoun{adrian09}.

\subsubsection{Interplay of Nonlinearity and Disorder and its Implications on
Anderson Localization.}

One of the particularly interesting recent developments on the
front of DNLS equations has been the examination of the
interplay of nonlinearity with disorder, in an effort
to explore how the presence of the former affects the
fundamental phenomenon of Anderson localization due to the
latter as first illustrated in \citeasnoun{anderson}.
This topic has spurred
a significant controversy within the physics and mathematics
communities, given its fundamental nature and the somewhat
conflicting results reported recently. In particular,
a number of recent experimental studies in optical
[see the works of \citeasnoun{moti} and
\citeasnoun{moti1}] and  atomic
[see e.g.~the work of \citeasnoun{prls1}]
physics settings have reported the observation of the
linearly induced exponential localization proposed by
Anderson for a disordered linear lattice. Nevertheless,
in a series of recent computational publications
in the physics literature by at least two separate
groups (whose theoretical arguments produce different results)
in \citeasnoun{flach08}, \citeasnoun{flach09}, and \citeasnoun{pikovsky08},
it has been argued that nonlinearity essentially {\it destroys}
Anderson localization by producing a
subdiffusive scaling of a quantity such as the second moment
%$M_2=\sum (i-i_0)^2 |\psi_i(t)|^2$ (where $|\psi_i(t)|^2$
%plays the role of the wavefunction density $\psi_i$ in
$M_2=\sum (n-n_0)^2 |u_n(t)|^2$ (where $|u_n(t)|^2$
plays the role of the wavefunction density in
the Schr{\"o}dinger problems considered therein). This
subdiffusive scaling implies that progressively
more distant lattice sites are occupied (although
slowly) and the process is not stopped by the
trapping anticipated on the basis of the
Anderson mechanism.
More specifically, it has recently been
argued in \citeasnoun{flach09} that there exist three regimes
within the system's spectral dynamics. For
weak nonlinearity (smaller than necessary
to induce transitions between linear modes), the system is
in a transient Anderson localization regime, but eventually it
detraps from it and grows according to the subdiffusive scaling
$M_2(t) \sim t^{\alpha}$, with $\alpha < 1$. For intermediate
nonlinearities, the subdiffusion of $M_2$ is initiated
immediately, while large nonlinearity
leads to localization of a fraction of the initial
wavepacket, while the ``radiative'' remainder is also subject
to this subdiffusive expansion. In the more recent work of
\citeasnoun{skokospre09}, the relevant considerations
were extended to more complex initial data (not only of
single site, but also of single mode or finite size), and
also a similar phenomenology was proposed in
\citeasnoun{khomerikiflach} to arise in the setting
of DNLS equations with linear (so-called Stark ladder)
potentials. In the latter, a discrete, equidistant spectrum
is also known to arise (but now due to the linear potential),
hence the same subdiffusion type dynamics is expected to ensue
in the presence of nonlinearity.
This controversy is extremely interesting, since it highlights the potential
``fragility'' of the Anderson localization regime, indicating that even
very weak nonlinearity is sufficient to eliminate the relevant phenomenology.

However, all the above considerations, at present, stay within the
realm of (admittedly, very strongly suggestive) numerical experiments.
Neither real experiments (e.g.,
with optical waveguides or BECs in optical lattices)
or rigorous mathematical arguments
have been provided to support this subdiffusional behavior.
It would be particularly interesting to try to examine such
issues from a rigorous point of view, also considering how
the relevant subdiffusive scalings may depend on the form (and distribution)
of the random
perturbations. Would similar phenomena also arise in the presence
of nonlinear random perturbations? Could the integrability of the
underlying nonlinear model crucially alter the dynamics (e.g., what
would happen to the above results if the DNLS was changed to the
AL-NLS model)?
These are only some of the numerous questions that this novel direction
raises.

\section{Klein Gordon Models in Mechanical and Electrical Systems}

\subsection{Recent Developments}

Interestingly, in recent years, there has been a parallel
stream of developments examining the dynamics of the localized
modes (discrete breathers ---DBs---) which are in some sense
the analogs of the standing
waves of the DNLS equation considered in the previous section.
Some of the fundamental tools that enabled the consideration
of the existence and stability theory of the discrete breathers
and multi-breathers were seeded in the original work of
\citeasnoun{ahnmacsep} and subsequently were generalized
e.g. in that of \citeasnoun{macsep}.
Similar results about the stability of such states had
been obtained earlier in \citeasnoun{arcetal}, using the
notion of the so-called Aubry band theory developed in \citeasnoun{aubry1},
however the results of the two methods can be shown to be
equivalent. In the presentation that follows, we will be
using the formulation of MacKay and collaborators, as was
adapted to the KG setting by \citeasnoun{koukpgk}. This has
the benefit that not only does it allow to quantify the
existence conditions of the relevant structures, but it also
offers tantalizing connections of the KG setting with the
previously explored DNLS one. It should also be noted that
in this setting, as well, the relevant results have not
been restrained to one-dimensional or just square lattices,
but have, in fact, been generalized to non-square lattices,
such as hexagonal and honeycomb ones in the works
of \citeasnoun{koukmac},
\citeasnoun{koukkour2} and \citeasnoun{kouketal}.

A short outline of the relevant methodology is as follows.
Assume that the system Hamiltonian is of the form
\begin{eqnarray}
H=H_0+\ep
H_1=\sum_{i=-\infty}^{\infty}\left(\frac{1}{2}p_i^2+V(x_i)\right)+\frac{\ep}{2}\sum_{i=-\infty}^{\infty}(x_{i+1}-x_i)^2,
\label{kg1}
\end{eqnarray}
where the part of $H_0$ is associated with an unperturbed
single oscillator, and $H_1$ denotes the perturbing part due
to the coupling with the neighboring oscillators.
Considering a number of central excited oscillators,
one can write their solution $x(t)$ into the action-angle
form $x(t)=\sum_{n=0}^{\infty} A_n(J) \cos(n w)$, where $J,w$
are the action-angle variables. Then, over the $n$ excited
oscillators, one can define the effective Hamiltonian
\begin{eqnarray}
H^{\mathrm{eff}}=H_0(I_i)+\ep\avh(\phi_i, I_i)\qquad i=1\ldots n \label{heff},
\end{eqnarray}
over the canonical variables:
\begin{equation}
\begin{array}{lll}
\vartheta=w_0 & &{\cal A}=J_0+\ldots+J_n\\
\phi_1=w_1-w_0& &I_1=J_1+\ldots+J_n\\
\phi_2=w_2-w_1& &I_2=J_2+\ldots+J_n\\
\vdots& &\vdots\\
\phi_{n}=w_n-w_{n-1}& &I_{n}=J_n\\
\end{array}\label{transformation}
\end{equation}
and with \[\avh=\frac{1}{T}\oint H_1 \ud t, \] where the integration is along
the unperturbed periodic orbit.

One of the key results of the theory of MacKay and collaborators
is that the critical points of this effective Hamiltonian
yield the DB solutions of the model. In the KG setting discussed
above, \citeasnoun{koukpgk} derived the effective Hamiltonian
in the form:
\begin{eqnarray}
\avh=-\frac{1}{2}\sum_{m=1}^{\infty}\sum_{i=1}^nA_m^2\cos(m\phi_i)\label{average}
\end{eqnarray}
and hence obtained the persistence conditions through
$\frac{\pa \avh}{\pa \phi_i}=0$. Remarkably, in this setting,
similarly to DNLS, it can again be proved that the only
available multi-pulse (i.e., multi-breather) solutions are
the ones with relative phase among the excited sites of
$0$ or $\pi$.

\begin{figure}[t]
\begin{center}
\includegraphics[width=4.5cm]{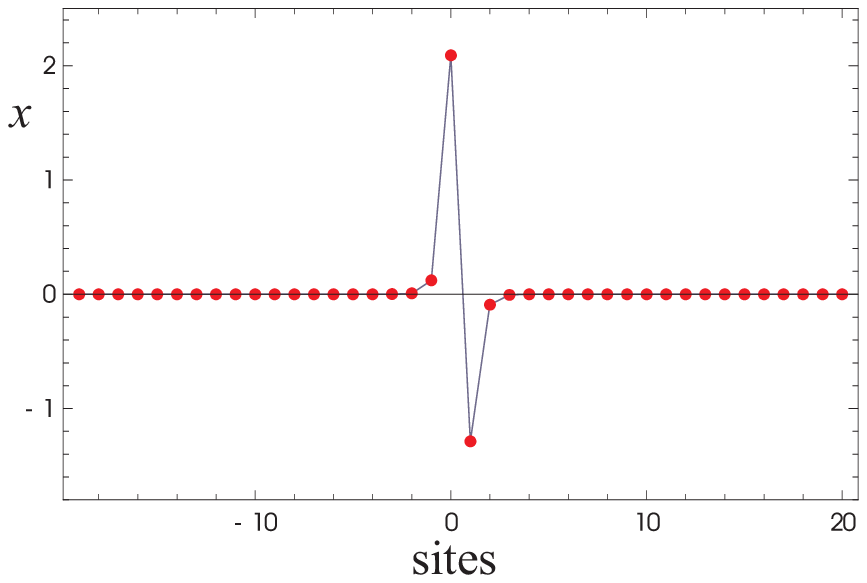}
~~
\includegraphics[width=4.5cm]{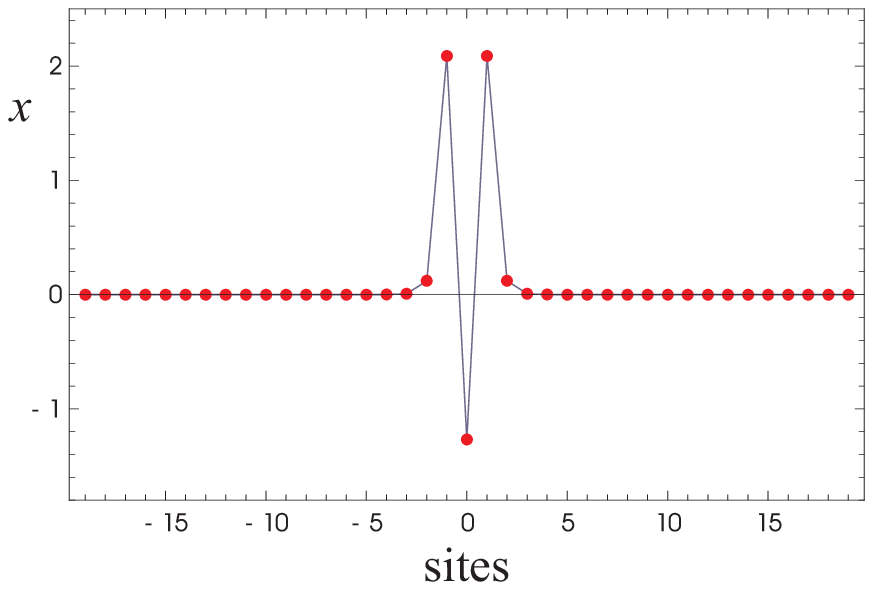}
~~
\includegraphics[width=4.5cm]{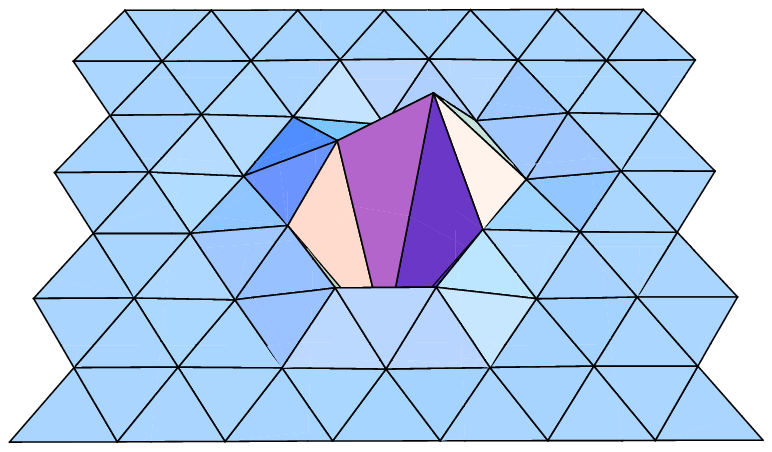}
\end{center}
%\begin{figure}[t]
%\begin{center}
%\includegraphics[width=4.5cm,height=5cm]{arc_review4.eps}
%\includegraphics[width=4.5cm,height=5cm]{arc_review5.eps}
%\includegraphics[width=4.5cm,height=5cm]{arc_review6.eps}
%\end{center}
%\vskip-0.4cm
\caption{(Color online) A sampler of breather solutions of the discrete nonlinear
Klein-Gordon equation in one and two spatial lattice dimensions with a (soft)
potential of the type used in the work of \citeasnoun{koukpgk}.
The left panel shows an out-of-phase two-site mode, while the middle
an out-of-phase three-site mode. The right panel shows a two-dimensional
hexagonal lattice generalization of these modes in the form of
a discrete vortex breather ($2 \pi/3$ phase difference between the
three adjacent excited oscillators), per the work of \citeasnoun{koukkour2}.
All three structures are stable close to the anti-continuum limit.}
\label{rev_fig7}
\end{figure}

Subsequently, one can examine the stability of these
DB solutions, which is controlled by the Hessian of
the effective Hamiltonian presented above. In particular,
${\bf E}={\bf J}
D^2H^{\mathrm{eff}}$ where the symplectic matrix reads
${\bf J}=\left(\begin{array}{cc}\bf O&-\bf I\\\bf I&\bf O\end{array}\right)$.
Once again, an explicit calculation of the relevant stability matrix
yields:
\begin{eqnarray}
%\fl
\bf E=\left(\begin{array}{c|c}\bf A&\bf B\\ \hline\bf C&\bf D\end{array}\right)=\left(\begin{array}{c|c}\ep\bf A_1&\ep\bf B_1\\ \hline\bf C_0+\ep\bf C_1&\ep\bf D_1\end{array}\right)=\left(\begin{array}{c|c}
-\ep\ds\frac{\pa^2\avh}{\pa\phi_i\pa I_j}&-\ep\ds\frac{\pa^2\avh}{\pa\phi_i\pa\phi_j}\\[10pt]
\hline\\[-8pt]
\ds\frac{\pa^2H_0}{\pa I_i I_j}+\ds\ep\frac{\pa^2\avh}{\pa I_i\pa I_j}&\ds\ep\frac{\pa^2\avh}{\pa\phi_j\pa I_i}
\end{array}\right).
\end{eqnarray}
For the special case of relative phases $\phi_i=0$, or $\phi_i= \pi$,
the matrix simplifies considerably acquiring the form:
\begin{eqnarray}
\bf E=\left(\begin{array}{cc}\bf O&\bf B\\\bf C&\bf O\end{array}\right)=\left(\begin{array}{cc}\bf O&\ep\bf B_1\\\bf C_0+\ep \bf C_1&\bf O\end{array}\right).\label{e1}\end{eqnarray}
which, subsequently, if we consider only the dominant eigenvalue
contributions (where $\lambda^2$ is of ${\cal O}(\ep)$), yields the relevant
squared eigenvalue matrix in the form:
\begin{eqnarray}
{\bf B_1}\cdot{\bf C_0}=-\frac{\pa\w}{\pa J}{\bf Z}=-\frac{\pa\w}{\pa J}\left(\begin{array}{ccccc}
2f_1&-f_1&0 & & \\
-f_2&2f_2&-f_2&0 & \\
 &\ddots&\ddots&\ddots & \\
 & 0 &-f_{n-1}&2f_{n-1}&-f_{n-1}\\
 &  &0 & -f_n&2f_n
\end{array}\right).\label{z}
\end{eqnarray}
In this expression $\omega=\partial H_0/\partial J$ denotes the
frequency, while $f_i=f(\phi_i)=(1/2) \sum_{n=1}^{\infty}
n^2 A_n^2 \cos(n \phi_i)$.
This leads to the characteristic exponents (i.e., effective eigenvalues)
of the DB in the form:
\begin{eqnarray}
\sigma_{\pm i}=\pm\sqrt{\ep\,\chi_{1i}}+{\cal O}(\ep^{3/2})\quad i=1\ldots n\label{sx},
\end{eqnarray}
where $\chi_{1i}=-\frac{\pa \w}{\pa J}z_i$ and $z_i$ are the eigenvalues
of the matrix ${\bf Z}$ defined above. Remarkably, the count of unstable
eigenvalues comes out to be identical to the DNLS case, for the
focusing or soft nonlinearities which satisfy $\ep \frac{\pa \w}{\pa J} < 0$.
That is, the number of unstable eigendirections will be precisely equal
to the number of nearest neighbors which are in phase between them.
Lastly, if the sign of the above product $\ep \frac{\pa \w}{\pa J}$
changes, then the conclusion indicates that the number of unstable
eigendirections is precisely equal to the number of out-of-phase
nearest neighbors. Some typical examples of the relevant configurations
in one- and two-dimensional settings
are shown in Fig. \ref{rev_fig7}.

%Some of the examples of the accuracy of these theoretical
%predictions for some typical two-site and three-site configurations
%(in particular, the in-phase ones, which should have one and
%two real eigenvalue pairs respectively) are shown in Fig.~\ref{rev_fig6}.

\subsection{Future Perspectives}
\label{future1}

While the general Hamiltonian framework of discrete breathers in
Klein-Gordon lattices is gradually being explored at a fundamental
level regarding the existence and stability properties that
it entails for its coherent structures, numerous new applications
are emerging in this field which require considerable additional
modeling efforts and which bear significant differences from this
general framework.

One such example is the recent work of \citeasnoun{lars},
where the dynamics of discrete breathers were examined
in arrays of coupled pendula, in the more realistic
experimental setting of forcing and damping. In the
Hamiltonian setting the relevant model for this system is simply the
discrete sine-Gordon equation, where the linear coupling stems
from the torsional springs, and the sinusoidal nonlinearity from
the gravitationally induced torsion. However, the actual
forced-damped array is more accurately described [as argued
in \citeasnoun{lars}] by:
\begin{eqnarray}
\ddot{u}_n + \omega_0^2 \sin(u_n) - \frac{\beta}{I} \Delta_2 u_n
= - \frac{\gamma}{I} \dot{u}_n  + \frac{\gamma_2}{I} \Delta_2 \dot{u}_n
- F \omega_d^2 \cos(\omega_d t) \cos(u_n),
\label{pendula}
\end{eqnarray}
where, in addition to the intrinsic frequency $\omega_0$ of each
pendulum and their torsional coupling of strength $\beta$,
terms for the damping of each pendulum (characterized by $\gamma$),
the viscous damping due to their coupling (characterized by $\gamma_2$)
and the strength $F$ of their parametric drive (characterized also by
the frequency $\omega_d$) have been included; $I$ denotes the pendulum's
moment of inertia. In this context, a number of interesting features
have been highlighted in the work of \citeasnoun{lars}, including
the apparent exchange of stability between the onsite (i.e., site-centered)
breather, and the inter-site (i.e., bond-centered) form. Both of
these waveforms were experimentally observed, each in their range
of stability, illustrating that there is a separatrix between the
respective ranges. The existence of such a separatrix connects partially
also to the discussion of the previous section about the Peierls-Nabarro
barriers, and their disappearance, as well as of the latter leading
to mobility of the localized modes. The mobility of
discrete breathers was, in fact, also observed in
\citeasnoun{lars} both experimentally and numerically.

 Nevertheless, there are still numerous topics within this theme
that remain partially unexplored in this more elaborate, yet more
realistic class of models. On the one hand, it would be useful
to explore some of the instabilities giving rise to coherent structures
therein, such as the modulational instability. Experiments already
show the presence of such phenomena as analyzed
by \citeasnoun{lars_private}, but
it would be useful if a quantitative description thereof could be
provided. Another interesting direction worth expanding towards
is that of multi-pulses. From the general theory of the previous subsection,
an understanding is emerging about which multi-pulse solutions (and under
what phase differences) should be expected to be stable/dynamically robust.
Developing a similar framework of understanding for the forced-damped
setting initially from a
computational perspective (and possibly also analytically at a later stage)
would certainly be desirable.

A similar driven damped setting has recently been explored in the context
of driven and damped electrical lattices in the work of
\citeasnoun{epl07} and of \citeasnoun{lars2}.  There
the experimental system is a bi-inductive discrete electrical
transmission line, which is nonlinear due to the nonlinear dependence of the
capacitance of the diode on the voltage and which
leads to the following dynamical model:
\begin{eqnarray}
\ddot{u}_n +  \left( \omega_0^2 u_n - \frac{\omega_0^2}{2 Q_0} u_n^2
+ \frac{\omega_0^2}{3 Q_0^2} u_n^3 \right) + \omega_1^2
\Delta_2 u_n &=& - \frac{\omega_1^2}{2 Q_0} \Delta_2 u_n^2
 + \frac{\omega_1^2}{3 Q_0^2} \Delta_2 u_n  ^3 + \frac{1}{R} \frac{d V_d}{dt}
\notag
\\[1.0ex]
&& + \frac{1}{R C_0} \left[1 - \left( \frac{u_n}{Q_0} \right) +
\left( \frac{u_n^2}{Q_0^2} \right) \right].
\label{electric}
\end{eqnarray}
Various electrical constants are denoted by their respective symbols
($R$ for resistance, $C_0$ for capacitance and $Q_0$ for charge),
while the external driver is given by $V_d$.
Some of the very interesting features observed in this system include
how its discrete breathers could interact with impurities (being e.g. seeded by
or attracted to a capacitor, while they could be destroyed or
repelled by inductors and/or resistors). Also, it was found
that depending on the driver amplitude and especially the driver
frequency, the system would spontaneously lock into 1- or 2- or
multi-breather states within a finite size (e.g. with 24 nodes)
experimental lattice. It would be extremely interesting
to try to explore the existence and the stability of localized
modes within such transmission line systems and to explore
the origin of the observed pattern selection and dynamics.
To the best of our knowledge, very little has been done theoretically
in this direction for such more realistic systems.

Interestingly, electrical lattices constitute, arguably, the
simplest realization of one of the focal points of interest
in the last few years (which is likely to be one of the
major areas of development for the next decade), namely of
nonlinear metamaterials~\citeasnoun{kozyrev}. 
Such media have simultaneously negative
electric permittivity and magnetic permeability (so that
their root square product, i.e., the speed of light propagation is
real); however, because of these negative values, they feature
a number of unusual effects, including negative refraction,
opposite signs of group and phase velocity (i.e., wavefronts
and wavepackets move in opposite directions), inverse Doppler
effect, backward Cerenkov radiation, etc. A nice introduction
to this field can be found e.g.~in the work of \citeasnoun{marques}
[see also for a review with a more nonlinear slant the presentation
of \citeasnoun{shalaev}]. One of the simplest (if not the
simplest) lattice that produces such backward waves can be generated
by feeding a ladder network with alternating shunt-connected
inductors and series capacitors [as shown e.g.~in Fig.~3.1 of
\citeasnoun{marques}]. However, recent work in the realm of
nonlinear waves has mostly focused on the slightly more elaborate
setting of magnetic metamaterials and, in particular, lattices
of split-ring resonators. The resonators are electrically equivalent
to an RLC oscillator, with self-inductance $L$, ohmic resistance $R$ and
capacitance $C$ and become nonlinear because of the gaps that they
possess which are filled with a nonlinear dielectric medium
(whose permittivity depends on the intensity of the electric
field $E$). The relevant mathematical description of the dynamics
of the charge stored on the capacitor of the $n$-th
split-ring resonator (SRR) of such a
SRR lattice, coupled to the current circulating through it can
be obtained as follows:
\begin{eqnarray}
\frac{d Q_n}{dt} &=& I_n
\label{tsi1}
\\
\frac{L dI_n}{dt} &=& - I R_n + M
\left(\frac{d I_{n+1}}{dt} + \frac{d I_{n-1}}{dt} \right)
- f(Q_n) + {\cal E}.
\label{tsi2}
\end{eqnarray}
Here, ${\cal E}$ is the electromotive force (i.e., external drive)
induced in each SRR due to the applied field, while $f(Q_n)$
bears the nonlinearity of the model since the field-dependent
permittivity gives rise to a field dependent capacitance, which
in turn leads to its nonlinear dependence on the charge. $f$ can
be reasonably approximated by an odd power series in $Q_n$, as
is illustrated e.g. in the work of \citeasnoun{lazaridis1}.
This formulation of, once again, a damped and driven electrical
lattice has been used in the work of \citeasnoun{lazaridis1}
to predict the existence of dissipative breathers
(as well as Hamiltonian ones, in the absence of the driving and
damping) under an AC (alternating current) 
electromotive force. This work initiated
a path towards a more systematic exploration of such structures,
which was continued in the work of \citeasnoun{lazaridis4},
examining such structures in both 1- and 2-dimensional settings,
that of \citeasnoun{maluckov} for different types of nonlinearities,
of \citeasnoun{lazaridis3} and \citeasnoun{lazaridis4}
for surface variants of these discrete
breathers and of \citeasnoun{molinalaz} for binary metamaterials.
More recently, it should be mentioned that further ``frequency gap''
regions in such metamaterials have been explored e.g. in the work of
\citeasnoun{tsitsas}, where the signs of the magnetic permeability
and dielectric permittivity are opposite; this regime has been
predicted to also possess breathers due to its approximate description by the
so-called short-pulse equation [see e.g., \citeasnoun{shaef}].
However, it should be mentioned that in most of these settings,
the relevant models are derived, and subsequently are analyzed
in a purely numerical form. Both a theoretical/mathematical
analysis of the types of structures that are possible,
and experimental realizations of the relevant settings are
presently lagging behind and would be a natural direction
for future work.

\section{FPU-type Models in Granular Crystals and Dusty Plasmas}

\subsection{Recent Developments \& Future Perspectives}

FPU models have a time-honored history of more than half a century,
one that has been captured in both special volumes
such as that of \citeasnoun{campbellchaos}, as well as in recent books
such as the one by \citeasnoun{galavas}. Despite the interesting
recent developments in this area, including the $q$-breathers of
\citeasnoun{flachq} (which may be closely related to the original
FPU paradox), here we will not focus on this intensely-studied
theme. Instead, we will present some particular recent directions
within the same general class of models which feature
predominantly intersite nonlinear interactions within a lattice
setting.

The principal example of the above class that we will consider
is that of the so-called ``granular crystals'', namely arrangements
of densely packed spheres, which upon suitable excitations interact
elastically through the so-called Hertzian contacts,
and result in the propagation of compression waves
(or, as we will see, the localization of suitable breather-like
states) in these media. These chains have
drawn considerable attention during the past few years.  This broad interest
has emerged, to a considerable extent, due to the wealth of available
material types/sizes and the ability to tune their dynamic response to
encompass linear, weakly nonlinear, and strongly nonlinear
regimes as shown in the works of e.g.~\citeasnoun{nesterenko1},
\citeasnoun{sen}, \citeasnoun{nesterenko2} and \citeasnoun{coste97}.
Such flexibility makes them perfect candidates
for many engineering applications, including shock and energy absorbing
layers proposed e.g.~in the works of \citeasnoun{dar06},
\citeasnoun{hong05}, \citeasnoun{fernando} and \citeasnoun{doney06};
actuating devices [see e.g.~\citeasnoun{dev08}]; and sound scramblers
[see e.g.~\citeasnoun{dar05} and \citeasnoun{dar05b}].

As an aside, we should note here that in an area such as that of
granular crystals, which from the
perspective of mathematical results is still at an early stage
(most of the focus has been on numerical investigations and
experimental results), it is not easy to distinguish between
recent developments and future perspectives. For this reason,
(and differently than in the previous sections)
we will present both of these in a unified format in what follows.

The model used to represent the simplest form of such
granular systems consists
of a one dimensional (1D) chain of spherical beads regulated by Hertzian
contact interaction potentials, see e.g.
\citeasnoun{Johnson}. For
this class of models, a
nonlinear wave theory, supporting nearly-compact solitary waves
was derived for all structured homogeneous [as discussed e.g. in
\citeasnoun{nesterenko1}]
and later also for heterogeneous periodic [as illustrated in
\citeasnoun{kd1}] materials
showing a highly nonlinear force ($F$)-displacement ($\delta$) response
dictated by the intrinsically nonlinear potential of interaction between
its fundamental components. These general nonlinear spring-type lattice
interactions read:
\begin{equation}
F = -V'(u_{i}-u_{i-1})=A\delta^p, \quad \delta=\max(0,u_{i-1}-u_{i}), \quad
\ddot{u}_i=-V'(u_{i}-u_{i-1})  + V'(u_{i+1}-u_{i})
\label{eq1}
\end{equation}
where $A$ depends on the material properties and $p$ is the nonlinear exponent
of the contact interaction (with $p>1$);
%Note that this
%force-displacement law is essentially nonlinear, that is nonlinearlizable;
%the lack of a linear part in this type of nonlinear stiffness laws has
%interesting implications on the dynamics \citeasnoun{Vakakis2008};
$p=3/2$ for spheres (Hertzian interactions).
By granular matter, we mean an aggregate of particles in
elastic contact with each other, preferably in linear or network shaped
arrangements as discussed e.g. in \citeasnoun{Goddard1990},
\citeasnoun{Gilles2001}, \citeasnoun{Goldenberg2005} and
\citeasnoun{Hostler2005}.
In addition to the nonlinear contact interaction, and related purely
to the particle geometry, another unusual feature of the granular state is
the zero tensile strength, which introduces an additional nonlinearity
(asymmetric potential) to the overall response. In the absence of static
precompression, these properties result in an essentially nonlinear force
which leads to interesting implications on the dynamics
%\citeasnoun{Vakakis2008}
%negligible linear range of the interaction forces between neighboring particles leading to a material with a characteristic
and a vanishing sound speed $c_0=0$ (so-called ``sonic vacuum''); see e.g.
\citeasnoun{Vakakis2008}.
This makes the linear and weakly nonlinear continuum approaches based on
the Korteweg-de Vries (KdV) equation  invalid and places granular
crystals in a
strongly nonlinear regime of wave dynamics. This feature uncovers a
significant
potential for relevant  applications. In particular, it supports a new
type of nearly
compact highly tunable solitary waves that have been experimentally
and numerically observed for 1D Hertzian granular systems
as summarized e.g.~in the reviews of \citeasnoun{nesterenko1} and
\citeasnoun{sen}.
%\citeasnoun{coste97}, \citeasnoun{Coste1999},
%\citeasnoun{Lazaridi1985,Sen1998,Sinkovits1995,coste97,Coste1999,Job2005,Vergara2005,Daraio2,Daraio3,Rosas2004,Rosas2007,Sokolow2007,kd3}.
While spherical bead crystals have been
mostly studied, similar responses can be obtained from cylindrical and
elliptical particles [see e.g.~\citeasnoun{Daraio09}], fibrous layers
[see e.g.~\citeasnoun{Lambert1984}] and  foams [see e.g.~\citeasnoun{sen}].

\begin{figure}[t]
\begin{center}
\includegraphics[width=6.75cm]{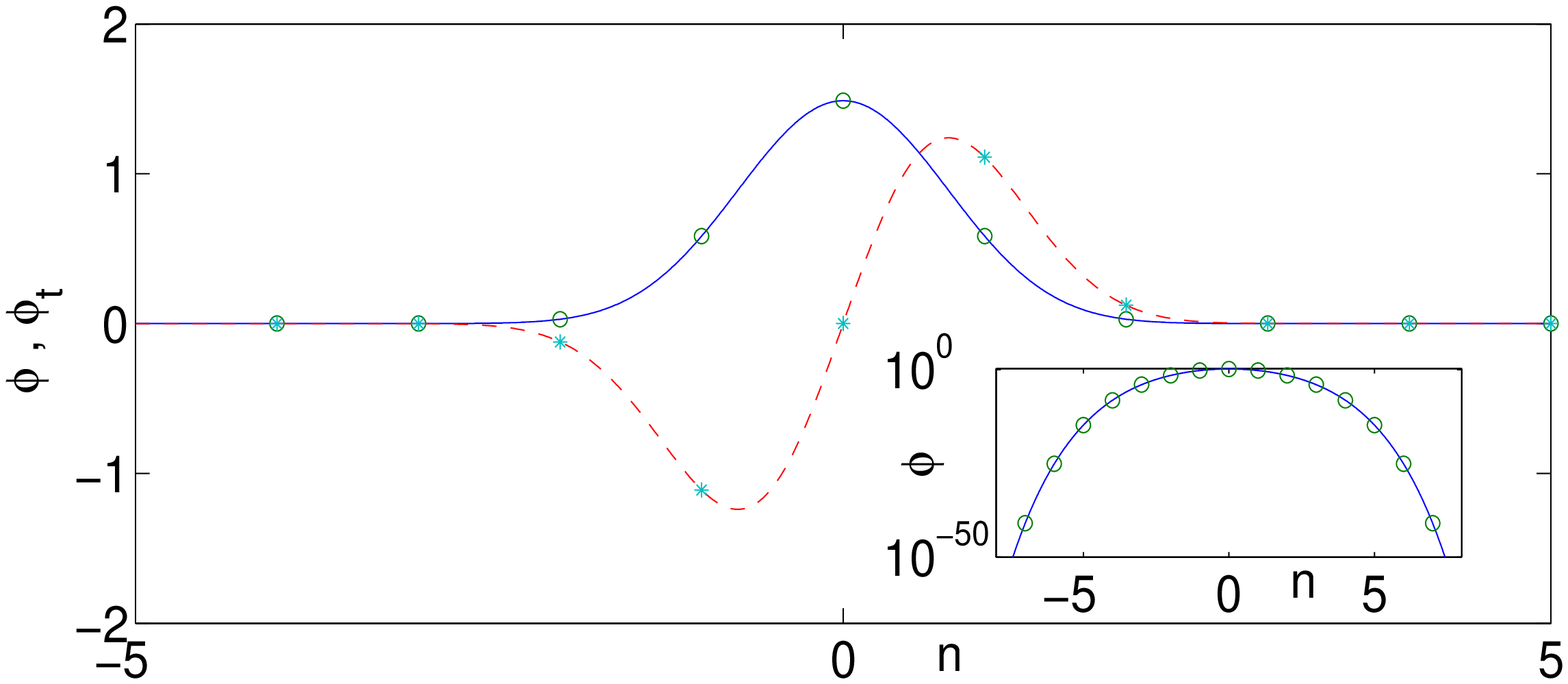}
~~
\includegraphics[width=6.75cm]{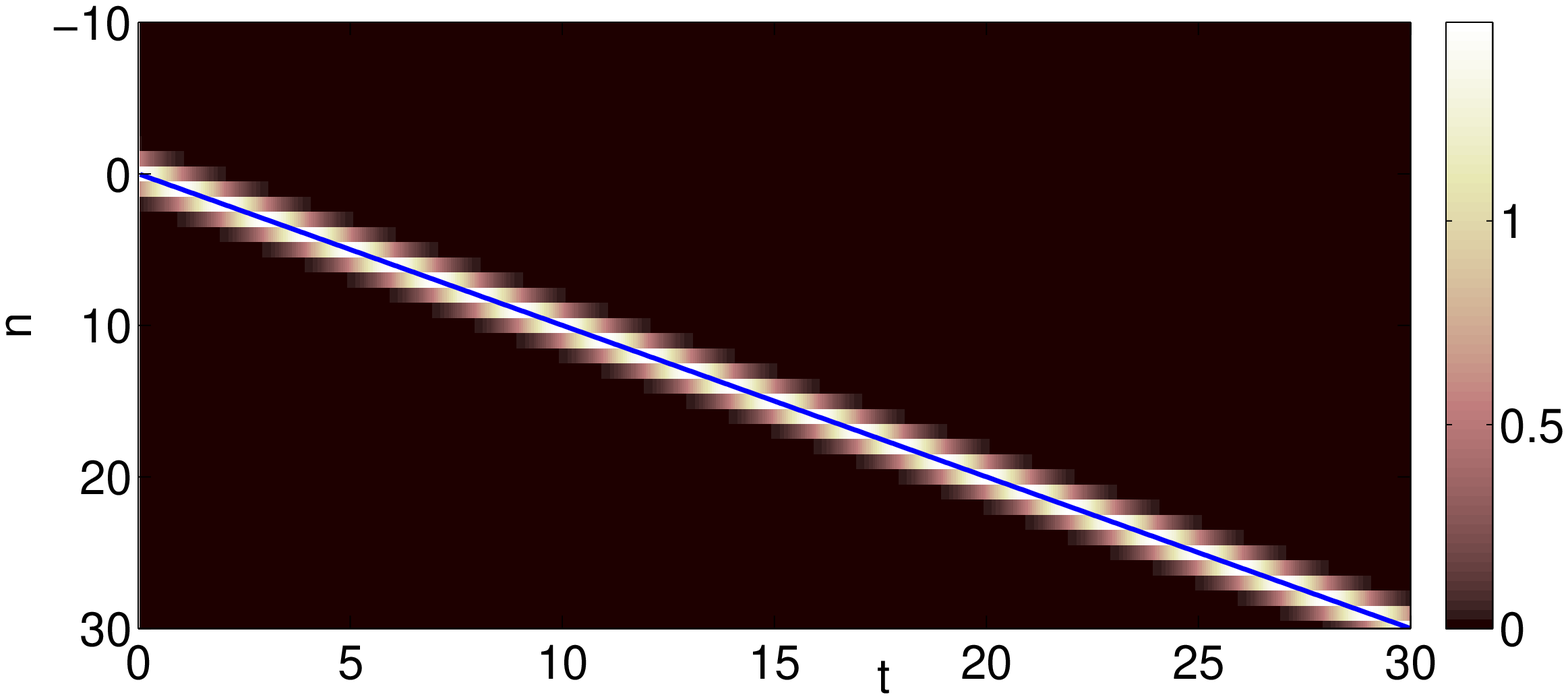}
\end{center}
\caption{(Color online) A typical example of the spatial profile of the positions
(blue solid line connecting the lattice nodes) and of the velocities
(red dashed line connecting the lattice nodes) of an exact
(nearly-compact, see inset)
traveling wave solution of the monomer granular chain, obtained through
the solution of Eq. (\ref{eqn2}). The space ($n$) - time ($t$) evolution of
this wave with its prescribed speed of unity is shown through the
contour map of the strain $r_n(t)$.}
\label{rev_fig8}
\end{figure}

The prototypical setup of Hertzian contacts describing elastic
interactions among grains of a unary (i.e., monoatomic) lattice
in the absence of any precompression (i.e., in the
so-called sonic-vacuum regime) can be written as:
\begin{eqnarray}
\ddot{u}_n = [u_{n-1}-u_n]_+^p - [u_n-u_{n+1}]_+^p.
\label{eqn1}
\end{eqnarray}
Here, the $[\cdot]_+$ notation indicates that
that the quantity in the bracket is only evaluated if positive,
while it is set to $0$, if negative.
%For the special Hertzian
%case of elastic grain interactions, $p=3/2$.
Reformulating the problem based on the strain variant of the
equation for $r_n=u_{n-1}-u_n$, we obtain
\begin{eqnarray}
\ddot{r}_n=[r_{n+1}]_+^p - 2 [r_n]_+^p + [r_{n-1}]_+^p.
\label{eqn2}
\end{eqnarray}
If we now seek traveling wave solutions, as suggested by the
above mentioned numerical and
experimental work (as well as the near-continuum asymptotics used) by
numerous authors [see e.g.~the presentations of \citeasnoun{nesterenko1}
and \citeasnoun{sen}], then $r_n=\phi(x) \equiv \phi(n-c t)$.
This leads to an advance-delay equation the solution of which,
from a rigorous viewpoint, ``remains an open question and apparently
a delicate one'' as indicated by \citeasnoun{pego}. This is because, although the latter
problem can be reformulated as an iterative scheme of the form:
\begin{eqnarray}
\phi_m(x)= \frac{(\Lambda \cdot \phi_{m-1}^p)(x)}
{\int_{-1}^1 (\Lambda \cdot \phi_{m-1}^p)(x) dx},
\label{eqn22}
\end{eqnarray}
with $\Lambda(x)=[1-|x|]_+$, the proof of existence of
a localized solution thereof for any $p>1$ is presently missing.
Nevertheless, from a numerical perspective, the scheme
of Eq. (\ref{eqn22}) has been shown to converge
to the desired solutions in the works of \citeasnoun{pego} and
\citeasnoun{pikovsky}; for an example of the relevant
traveling wave in the Hertzian case, see Fig. \ref{rev_fig8}.
It should also be
mentioned that
should such a proof be available, it would immediately
imply that
\begin{eqnarray}
r(x+1) \leq {\rm sup}_{y \in [-1,1]} r^p(x+1-y) \int_{-1}^1 \Lambda(x)
=r^p(x),
\label{eqn23}
\end{eqnarray}
which, in turn, suggests the double exponential law
$r(x+n) \leq r(x)^{p^n}$ for the decay of the solution tail.
It is this extremely fast decay which leads the continuum
analog of the solution to sustain compactly supported structures.
However, it should be noted that this latter statement has not been rigorously
proven either (to the best of our knowledge). In fact, there
are multiple continuum models that have been proposed
for the system dynamics. The most standard of them
stems from the long wavelength approximation (LWA) applied to the
displacement problem by \citeasnoun{nesterenko1} [see also
\citeasnoun{coste97}], while
a more recent approach in the work of \citeasnoun{pikovsky}
suggests that the LWA be applied to the equation for
the strains. The former approach leads to the partial
differential equation (PDE), for the
strain variables, of the form:
 \begin{eqnarray}
r_{tt}=(r^p)_{xx} + \frac{\epsilon^2}{12} \left( (r^p)_{xxxx} +
\frac{n (n-1)}{2} (r^{p-2} r_x^2)_{xx} \right),
\label{eqn26}
\end{eqnarray}
while the latter establishes:
\begin{eqnarray}
r_{tt}=(r^p)_{xx} + \frac{\epsilon^2}{12} (r^p)_{xxxx}.
\label{eqn27}
\end{eqnarray}
In fact, from a mathematical perspective, one can go even
further and suggest that since the above models are singularly
perturbed (and may suffer relevant pathologies, which may also affect
their computational implementation), it may be relevant
to suggest the consideration of regularized variants of
these PDEs, such as the one proposed by
the work of \citeasnoun{rosenau1} and \citeasnoun{rosenau2} as:
\begin{eqnarray}
r_{tt}= (r^p)_{xx} + \frac{\epsilon^2}{12} r_{xxtt}.
\label{eqn28}
\end{eqnarray}
A systematic examination of the relevant models and a rigorous
proof of the decay properties of its associated solutions
(and how they connect to those of the discrete problem from which
the approximation stems)
would be a particularly interesting theme for future
studies.
%
%\subsection{Future Perspectives}
%In an area such as that of granular crystals, which from the
%perspective of mathematical results is still at an early stage
%(most of the focus has been on numerical investigations and
%experimental results), it is not easy to distinguish where
%recent developments end and future perspectives

A recent direction of intense interest within this theme
of granular crystals has concerned the examination of
progressively more heterogeneous settings.
In the work of \citeasnoun{kd1} and \citeasnoun{kd2},
dimers of different types were considered (e.g.,
1 steel bead-1 teflon bead, or 2 steel-1 teflon, $\dots$, or 5 steel-1
teflon etc., as well as similarly with other materials such as brass,
rubber, glass and nylon), as well as of trimers (1 steel bead, 1 brass
bead, 1 teflon bead or 1 steel, 1 glass, 1 nylon, etc.).
From a theoretical/mathematical perspective, one of the
fundamental contributions of this work for the consideration of
\begin{align}
m_1 \ddot{u}_j &= (w_j-u_j)^k-(u_j-w_{j-1})^k\,, \label{teq1} \\
m_2 \ddot{w}_j &= (u_{j+1}-w_j)^k - (w_j-u_j)^k\,, \label{teq2}
\end{align}
(i.e., the reduced dimer granular setting) was the proposition of
a modified Taylor expansion:
\begin{align}
	w =\lambda \left(u + b_1 D u_x + b_2 D^2 u_{xx} + b_3 D^3 u_{xxx}
%\right. \notag \\
%		&\quad \left.
+ b_4 D^4 u_{4x}  + \dots \right)\,. \label{eq3}
\end{align}
This enabled the effective ``homogenization'' of the ordinary
differential equations into a ``dimer continuum medium'', provided
that the relevant consistency [between Eqs.~(\ref{teq1}) and (\ref{teq2})]
conditions $\lambda=1$, $b_1=1$, $b_2=m_1/(m_1+m_2)$,
$b_3=(2 m_1-m_2)/(3 (m_1+m_2))$, $b_4=m_1 (m_1^2-m_1 m_2 + m_2^2) /
(3 (m_1+m_2)^3)$, etc. are satisfied. Notice that for $m_1=m_2$,
these yield the regular  Taylor expansion, as expected.
This long-wavelength approach for the dimer problem produced a
PDE of the type of Eq. (\ref{eqn26}), whose solutions were
compared favorably to numerical computations and experimental
results in the works of \citeasnoun{kd1} and \citeasnoun{kd2}.

On the other hand, the consideration of dimers enabled for the first
time another possibility that has been very recently explored more
systematically. In particular, in the presence of precompression, the
dimer model becomes:
\begin{eqnarray}
	m_{i}\ddot{u}_i = A[\delta_{0}+u_{i-1} - u_{i}]^{3/2} - A[\delta_{0}+u_{i} - u_{i+1}]^{3/2},
%\\
%	A &=& \frac{2E\left(\frac{R_iR_{i+1}}{R_i + R_{i+1}}\right)^{1/2}}{3\left(1-\nu^2\right)}\,,	
	\label{model}
\end{eqnarray}
where 	
%$A = \frac{2E\left(\frac{R_iR_{i+1}}{R_i + R_{i+1}}\right)^{1/2}}{3\left(1-\nu^2\right)}$,
$m_{2i+1}=M$ and $m_{2i}=m$ for all $i \in Z$ (for same material,
different radii $R_i$ beads in the simplest realization).
Then,
at the end of the first Brillouin zone, i.e., at $k=\frac{\pi}{2\alpha}$, where $\alpha=R_{i}+R_{i+1}-\delta_{0}$ is the equilibrium
distance between two adjacent beads, the linear spectrum
possesses a gap between the upper
cutoff of the acoustic branch, $\omega_{1}=\sqrt{2K_2/M}$,
and the lower cutoff of the optical one,
$\omega_{2}=\sqrt{2K_2/m}$,
where $K_2=\frac{3}{2}A\delta_{0}^{1/2}$. This gap and the corresponding
reformulation
of the granular chain as an approximate $K_2-K_3-K_4$ chain of the
form:
\begin{eqnarray}
m_{i}\ddot{u}_i &=&  K_{2}(u_{i+1}-2u_{i}+u_{i-1})
+K_{3}\left((u_{i+1}-u_{i})^2-(u_{i-1}-u_{i})^2 \right) \nonumber \\
&+& K_{4}\left((u_{i+1}-u_{i})^3+(u_{i-1}-u_{i})^3 \right)
\label{K2K3K4}
\end{eqnarray}
 with $K_3=-\frac{3}{8}A\delta_{0}^{-1/2}$ and
$K_4=\frac{3}{48}A\delta_{0}^{-3/2}$ enables the adaptation of the
findings of \citeasnoun{prb}, demonstrating that since
$\frac{K_3^2}{K_2K_4}>\frac{3}{4}$,
discrete gap breather modes will bifurcate from the optical band,
in fact, due to the {\it modulational instability} of
the band edge. The relevant nonlinear states were recently
experimentally and numerically established in the work of
\citeasnoun{boechler}. Furthermore, such breathing excitations
were also established in the presence of impurities by
\citeasnoun{kavous}, as a nonlinear continuation of the linear
modes induced by the presence of such defects. In that setting
the relevant nonlinear impurity modes were obtained both
in the setting of one defect, as well as in that of a defect
``double well'' (with two defects separated by one lattice node).
Similar excitations but of a transient form, due to the absence
of precompression (and hence of an underlying linear regime), were
also  recently observed experimentally (and discussed theoretically)
in the work of \citeasnoun{job2}.

It is interesting to also note that once the role of individual
defects is starting to be explored, it is natural to examine the
progressive inclusion of disorder within such granular chains.
A few steps have been already taken in that direction.
In particular, in the work of \citeasnoun{ponson}, a systematic
approach towards increasing randomness within the lattice has
been used, by examining a dimer setting, and progressively
``flipping'' the order of the masses in the dimer in order
to create defects. In that case, a transition between the
propagation of waves in low-disorder regimes and the degradation
of the wave in high-disorder lattices was clearly illustrated.
On the other hand, the incorporation of disorder can be used
to achieve a degree of optimality in granular networks. An
example in this direction is the work of \citeasnoun{fernando},
where upon a given selection of masses and elastic properties
thereof, a genetic algorithm approach was used towards the
optimization (minimization) of the propagated force at the
end of the chain. This was a first step towards the optimal
construction of ``granular protectors''.

The above investigations are indicative of the intense
recent activity in the field of granular crystals. Yet, there
is a wide range of topics that require additional theoretical,
numerical and experimental investigation. In the one-dimensional
setting, in addition to exploring the existence (and stability)
of traveling waves from a rigorous viewpoint not only in the
monomer, but also in the ``multimer'' (i.e., dimer, trimer etc.)
setting, it would also be of interest to explore more systematically
the role of disorder. In particular, the setting of granular chains
might be an ideal testbed where the recent developments of
subdiffusive wavepacket spreading and possible ``destruction'' of Anderson
localization in nonlinear lattices could be explored controllably. This is because
this setting can be made, on demand, nearly linear, weakly or strongly
(or even completely) nonlinear. Furthermore, while the above
ideas discussed earlier based on the works of
\citeasnoun{flach08}, \citeasnoun{flach09},
\citeasnoun{pikovsky08} and \citeasnoun{skokospre09}, were
established in the context of nonlinear models with onsite
potentials such as the DNLS, it would be interesting to explore
them in models within the FPU class i.e., with inter-site interactions,
and especially ones such as the granular chain which could in principle
be realized experimentally in a controllable fashion.
On the other hand, another important observation in many of
the above mentioned experimental works in granular chains
[see also the relevant reviews of \citeasnoun{nesterenko1} and
\citeasnoun{sen}] is the key role of dissipative perturbations
in decreasing the amplitude of the propagating [or localized
breathing as in \citeasnoun{boechler}] waves. A widely applicable,
quantitatively accurate modeling from first principles of such
mechanisms is still missing despite relevant recent efforts
and interesting corresponding results e.g. in the works of
\citeasnoun{rosasnes} and of \citeasnoun{ricardochiara}.
Another direction which still remains relatively pristine
is that of higher dimensional structures. While some of
the linear or near-linear behavior has been explored
therein in both ordered and disordered granular chains,
much less is known about the structure and propagation
(as well as the localization) of nonlinear waves within
such higher-dimensional granular crystals. This is a theme which, considering
the activity and progress recently spurred in the one-dimensional
setting, is likely to emerge as one of the most lively
areas of investigation within the next decade.

Finally, we wanted to add a brief comment on FPU-type lattice
applications in dusty plasmas and the potential emergence of
nonlinear excitations in these settings. In recent works
such as those of \citeasnoun{kourakis1}, \citeasnoun{koukkour2}
and \citeasnoun{kourakis3}, the setting of dusty plasmas has
been proposed as a canonical realization of nonlinear Klein-Gordon
lattices. The coupling between adjacent nodes/dust grains has been
proposed to be linear, while the sheath environment has been
argued to create a nonlinear onsite potential landscape that
drives the relevant transverse dynamics considered in these
works. In this context, both one- and two-dimensional 
discrete breathers
of the type explored in the previous section for Klein-Gordon
lattices have been identified and their stability has been
accordingly examined. However, here, we would like to focus
the attention instead on the broader context of e.g. the
presentation of \citeasnoun{kourakis3}. There both longitudinal
and transverse vibrations of the relevant dust particles were
examined. Although the latter are, in principle, experimentally
possible as in the work of \citeasnoun{melzer}, there is also
a particular interest in the former, as they incorporate the
nonlinear potential of interaction across sites which can be
accurately approximated by an exponentially dependent on the
inter-particle separation Yukawa-type landscape. Both the
linear and the nonlinear implications of such lattices have
been explored in some detail as e.g. in the recent review
of \citeasnoun{melzer1}. However, the emergence of nonlinear waveforms
and especially the existence, stability and mobility of discrete
breathers in these intersite but non-nearest-neighbor (in fact,
the coupling is ``all-to-all'') lattices have received limited
attention. We feel that this field is also presently at a ripe
phase, where the use of the tools and technology of localized
modes and their stability and dynamics could have a substantial impact in
experimental, as well as numerical investigations.

\section{Instead of an Epilogue}

The field of nonlinear dynamical lattices has seen a tremendous explosion
of interest associated with it over the past 20 years. Although the
first anharmonic lattice investigations were initiated more than 50 years
ago by Fermi, Pasta and Ulam, and part of these led to the remarkable
theory of solitary waves and solitons which has been pervasive of
various areas of science for more than the past 40 years, the appreciation
of the idiosyncrasies of the lattices in comparison to their continuum
siblings is far more recent. The seminal studies on intrinsically localized
modes 
%by some of the relevant pioneers such as Ablowitz, Dolgov, Sievers,
%Takeno, Page, Eilbeck, Scott, Lomdahl, Peyrard, Campbell, Kivshar, Flach,
%Aubry, MacKay, Bishop, Malomed, Remoissenet, Konotop, Salerno, Tsironis and
%the many that  followed was 
led to remarkable breakthroughs that fueled research in
areas as diverse as nonlinear optics and optical waveguides, Bose-Einstein
condensates in optical lattices, micromechanical systems, pendulum
arrays, electrical lattices, DNA denaturation, Josephson junction arrays,
split ring resonators and metamaterials, granular crystals, dusty
plasmas, metallic uranium and ionic sodium iodide, and possibly even
in sheets of graphene [see e.g., the recent work of \citeasnoun{yamayose}].
Alongside the development of the areas of application, the
particularities of each of these settings have forced not only a
development of the general mathematical ideas and of fundamental
conditions under which the relevant excitations should be expected
to exist and be robustly preserved by the system dynamics, but 
have also inspired the more specific adaptation of these ideas and
of the relevant modeling
to the special settings of each particular application.  They have
also motivated the emergence of a diverse array of
relevant excitations, ranging from solitons to 
discrete breathers, to
multi-peaked structures or even to compactly (or nearly compactly) supported
ones. It is clear that after the initial decade of enthusiasm
and exploration, and the second one of broad and far reaching
applicability, the field is now entering a new decade of maturity,
but where significant challenges lie ahead. These concern the
application of the relevant ideas to very recently emerging fields
(materials and metamaterials, granular media, plasmas),
their exploration in more realistic and especially higher dimensional
settings and the establishment of more firm connections/interplays with other
areas of physics, such as non-equilibrium thermodynamics, or the study
of disordered media and of wave phenomena within them. What the
resulting 20/20 vision of nonlinear waves in lattices will be
remains to be seen$\dots$

{\bf Acknowledgments}. This review would not have been possible without
the help, collaboration and insights of numerous colleagues that have
shaped my vision of the field through our interactions. As a partial
list, Dimitri Frantzeskakis, Ricardo Carretero (who also contributed
significantly towards improving this manuscript and is especially
thanked for that), Boris Malomed, Yuri
Kivshar, Volodya Konotop, Augusto Smerzi, Mario Salerno, Mason Porter,
Chiara Daraio, Yannis Kourakis, Vassilis Koukouloyannis, George Theocharis,
Jesus Cuevas, Dmitry Pelinovsky, Faustino Palmero, Wieslaw Krolikowski,
Michael Weinstein, Chris Jones, Todd Kapitula, Shozo Takeno, Al Sievers,
Alex Kovalev, Atanas Stefanov, Sergey Dmitriev, Alan Bishop, Avadh
Saxena, Avinash Khare, Kim Rasmussen, Zhigang Chen, Kody Law,
Andrea Trombettoni, Yuri Gaididei are especially thanked.
The support of the National Science Foundation through the
CAREER program (DMS-0349023) and through NSF-DMS-0806762, as well
as that of the Alexander von Humboldt Foundation through its research
fellowship program are gratefully acknowledged.

%%%%%%%%%%%%%%%%%bibliography style

\end{document}